\documentclass[twocolumn]{aastex631}
\usepackage{amsmath}
\usepackage{units}
\usepackage{listings}
\usepackage{wasysym}
\usepackage{natbib}
\usepackage{bm}
\usepackage{colortbl}
\usepackage{lipsum}
\definecolor{LightCyan}{rgb}{0.88,1,1}

\NewDocumentCommand{\codeword}{v}{%
\texttt{\textcolor{blue}{#1}}%
}

\usepackage{xcolor}
\usepackage{hyperref}
\begin{document}

\title{A pipeline for identifying astrophysical signals and instrumental glitches in LISA data}

\author[0009-0002-5780-4539]{Aryamann Rao}
\affiliation{Department of Physics and Astronomy, University of British Columbia, 6224 Agricultural Road,
Vancouver, BC V6T 1Z1, Canada}

\author[0000-0003-0316-1355]{Jess McIver}
\affiliation{Department of Physics and Astronomy, University of British Columbia, 6224 Agricultural Road,
Vancouver, BC V6T 1Z1, Canada}

\author[0000-0001-6877-3278]{Rhiannon Udall}
\affiliation{Department of Physics and Astronomy, University of British Columbia, 6224 Agricultural Road,
Vancouver, BC V6T 1Z1, Canada}

\author[0000-0002-1112-8830]{Natalia Korsakova}
\affiliation{Universite Côte d’Azur, Observatoire de la Côte d’Azur, CNRS, Artemis, Bd de l’Observatoire,
F-06304 Nice, France}

\author[0009-0003-1050-9273]{Malvina Bellotti}
\affiliation{Universite Côte d’Azur, Observatoire de la Côte d’Azur, CNRS, Artemis, Bd de l’Observatoire,
F-06304 Nice, France}

\begin{abstract} 
The Laser Interferometer Space Antenna (LISA) is a space-based gravitational wave detector scheduled for launch in the 2030s by the European Space Agency. Operating in the 0.1 mHz to 0.1 Hz band, LISA will observe a variety of astrophysical sources including but not limited to supermassive black hole mergers, extreme mass ratio inspirals, and white dwarf binaries. LISA will yield new insights, including supermassive black hole formation, new tests of general relativity, population studies of white dwarfs and others. Like all interferometric gravitational wave detectors, LISA data is expected to contain non-astrophysical transients, commonly referred to as glitches. We introduce a two step pipeline that consists of a search algorithm for detecting potential transients in simulated LISA data followed by a machine learning model that classifies the detected transients as gravitational waves or instrumental glitches. This is important for MBHBs with very high masses (around $10^6\ M_{\odot}$) which are transients that could be morphologically similar to LISA glitches in the data. Initial results demonstrate a promising ability to classify detected events with an accuracy close to $98\%$. The pipeline also shows the ability to analyze a month of data in under 2 minutes on a laptop. This shows an eventual utility of our pipeline for enabling accurate rapid alerts from low latency searches. It could also support model selection in global-fit analyses by providing an initial summary of the location and type of individual signals. This paper also discusses challenges encountered during training, including limitations of the current framework.
\end{abstract}

\section{Introduction}\label{sec:introduction}
Since 2015, ground-based interferometric gravitational wave (GW) observatories, such as those run by the LIGO-Virgo-KAGRA (LVK) collaboration, have been highly successful, having detected hundreds of GW signals from stellar-mass compact binary coalescences \citep{gwtc_1, gwtc_2, gwtc_3, gwtc_4}. The European Space Agency (ESA) and NASA are now preparing to launch a next-generation, space-based GW observatory known as the Laser Interferometer Space Antenna (LISA) in the 2030s \citep{lisa}, which will open a whole new window into the GW universe. Unlike ground based detectors which primarily detect GWs in the 10 Hz to 1~kHz frequency range \citep{ligo_bbh}, LISA will be sensitive to much longer lasting signals with frequencies ranging from 0.1 mHz to 0.1 Hz \citep{aet_tdi, lisa}. In this domain, LISA will observe new astrophysical sources invisible to current ground-based detectors such as Massive Black Hole Binaries (MBHBs) \citep{lisa_smbhb}, Extreme Mass-Ratio Inspirals (EMRIs) \citep{lisa_gr}, a galactic Double White Dwarf (DWD) inspiral background, \citep{lisa_bwd, lisa_bwd2} and compact binaries part of hierarchical systems containing multiple stars \citep{lisa_ht}. It will also observe resolved galactic binaries that does not form the background and bursts and backgrounds of cosmological origin.

However, like any interferometric GW detector, LISA will be susceptible to glitches: transient signals in the data stream, not caused by astrophysical sources (\autoref{subsec:lisa_pathfinder} and the citations within will explore this in further detail). These can pose a significant hurdle to downstream data-analysis for any GW detector, both ground \citep{ligo_glitch} and space-based \citep{lpf_glitches}. They interfere with GW parameter estimation, can overpower GWs in brightness and could mock the morphology of GWs. Therefore, accurate detection and mitigation of these instrumental transients is important to enable accurate GW alerts, skymaps and catalogs. 

\begin{figure*}
    \centering
\includegraphics[width=0.9\textwidth]{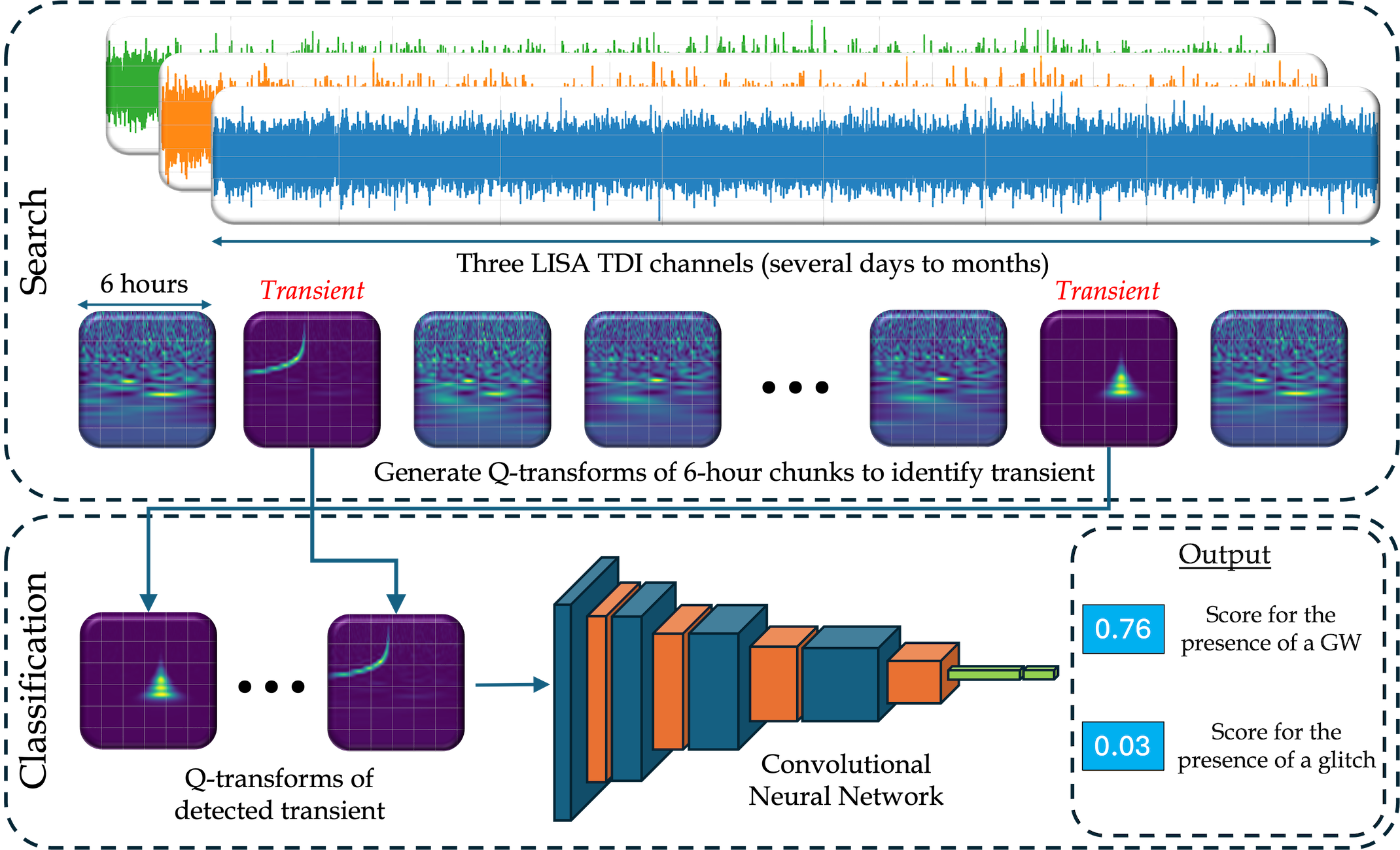}
    \caption{A flowchart describing our proposed pipeline for transient search and classification for LISA. The pipeline consists of two parts: a search algorithm and a machine learning classifier. The former takes as input, the LISA TDI channels (\autoref{subsec:pipeline}). It divides this data into 6 hour segments and generates Q-transforms (\autoref{subsec:q-transform}) of each. Any excess of energy is classified as a transient (\autoref{subsec:search_algorithm}) and the Q-transforms containing transients are passed on to the classifier (\autoref{subsec:architecture}).The classifier is a CNN that processes each input image and outputs two scores: its confidence that a GW is present and its confidence that a glitch is present. The scores each range from 0 to 1 with a higher number indicating a greater likelihood of the corresponding transient type being present in the image.}
    \label{fig:pipeline}
\end{figure*}

In this paper, we present a pipeline consisting of a search algorithm to detect transient GW and glitch events in LISA output data and a Convolutional Neural Network (CNN) for classifying identified transients as glitches and GWs from MBHBs. We use the word \textit{transient} to define any signal that appears temporarily in the data. Hence our usage of the word is not limited to short duration events. Some transients included in our work might persist for a longer period before disappearing. We also show that this approach works very well when GW and glitch transients are present in close proximity. We develop and test our pipeline on simulated LISA data. We structure our paper as follows. The following subsections will outline the fundamentals our work is built on. \autoref{subsec:lisa_pathfinder} discusses the LISA Pathfinder mission, which allowed for accurate modeling of glitches expected in the full-scale LISA mission. We use these results while simulating glitches for our study. \autoref{subsec:pipeline} gives an overview of our approach and compares it to previous work done in this area. \autoref{sec:dataset_generation} details the methods we used to generate LISA simulations which were required for our analysis. \autoref{sec:search_algorithm} describes the burst search algorithm which forms the first part of our pipeline. This is followed by \autoref{sec:classifier_training} and \autoref{sec:classifier_performance}, which outline how the classifier was trained and how it performed, respectively. We present the current limitations of our work, future avenues of research, and conclusions in \autoref{sec:discussion}.

\newpage
\subsection{LISA Pathfinder}\label{subsec:lisa_pathfinder}
In 2015, ESA launched the LISA Pathfinder (LPF) mission which served as a technology demonstrator for feasibility of laser interferometry in space. The payload consisted of one spacecraft, with a similar model to that of the component spacecraft on the final detector. The spacecraft housed two freely falling test masses with a system of lasers to measure the change in relative displacement between the test masses via laser interferometry. The primary objective was to test whether it is possible to maintain test masses in free-fall, with the amplitude spectral density of the residual acceleration noise being no greater than $30\ \text{fm s}^{-2}/\sqrt{\text{Hz}}$ in a frequency range of $1$ mHz to $30$ mHz \citep{pathfinder_results}. The resulting LPF data showed an ASD of the residual acceleration noise being $5.2\pm 0.1\ \text{fm s}^{-2}/\sqrt{\text{Hz}}$ between 0.7 and 20 mHz, thus demonstrating the feasibility of the new required technology \citep{pathfinder_overview}. However, LPF also detected non-astrophysical transient disturbances to the free fall motion of the test masses, which are now known as glitches. While the specific cause for these events is mostly unknown, studies have attempted to model these glitches with various phenomenological functions \citep{lpf_glitches}. \autoref{subsec:glitch_model} will discuss the LPF glitch models we used, in more detail.

\subsection{Our proposed pipeline}\label{subsec:pipeline}

The main LISA output will be beat note fluctuations from the spacecraft interferometers. This raw data is dominated by laser noise which can be eliminated using time delay interferometry (TDI). This involves combining the different readouts such that the laser noise becomes lower than the secondary noises (a comprehensive description is outlined in \citealp{tdi}). We can form three different channels from the raw LISA measurements. A variety of such combinations resulting in different TDI channels are available. One of the most commonly used are the second generation Michelson $X_2,Y_2,Z_2$ channels and their linear combinations
\begin{equation}
    A=\dfrac{Z_2-X_2}{\sqrt{2}},\
    E=\dfrac{X_2-2Y_2+Z_2}{\sqrt{6}},\
    T=\dfrac{X_2+Y_2+Z_2}{\sqrt{3}}.
    \label{eqn:tdi_channels}
\end{equation}

We produce a class of time–frequency representations for the LISA TDI channels, known as Q-transforms (\autoref{subsec:q-transform}), to support both the identification and classification of GWs and glitches. First, the timeseries representations of LISA data are divided into 6-hour segments, and a Q-transform is computed for each segment. In this study we will focus on the $A,E$ channels. This is because the $A,E$ combinations show reduced noise correlations as compared to the second-generation Michelson channels $X, Y, Z$ \citep{aet_tdi}. Adjacent Q-transform segments overlap by 1 hour to retain sensitivity to transients that occur at the boundaries. Then, an energy excess search algorithm similar to what is presented in \citep{alan_search} (and explored further in \autoref{sec:search_algorithm}) is used to detect the times at which potentially interesting transient events occur. The Q-transform of data containing detected transients are then input to a CNN to identify the transient type. This allows downstream analysis to greatly reduce the parameter space for finding GW sources and apply glitch mitigation early in the filtering process, increasing accuracy and decreasing computation time. This allows us to perform the models selection between astrophysical signals and glitches and process them with the corresponding parts of the pipeline without confusing them with each other. It also acts like a search for glitch signals (and any other potential burst signals because it is not model dependent). A flowchart of this pipeline is shown in \autoref{fig:pipeline}.

CNNs are among the most widely used ML models for transient classification. For instance, the LVK collaboration has used CNN models such as GravitySpy \citep{gravityspy} and GSpyNetTree \citep{gspynettree} which input Q-transforms of data from ground based detectors to classify detected transients. Our work will draw elements from both these studies. There have also been ML models applied to LISA data. For instance, \cite{lisa_cnn_2024} proposed a multi-stage approach using CNNs and Long-Short Term Memory (LSTM) to detect, classify, and subtract GWs and glitches. This framework was recently extended in \cite{lisa_cnn_2025} to explore direct, single-step source separation for overlapping GWs and glitches using encoders and decoders. However, our pipeline differs from both prior works in three key respects. First, while \cite{lisa_cnn_2024} modeled their transients using Fast Rise Exponential Decay (FRED) models, our framework uses physically motivated PyCBC waveforms for GWs (\autoref{subsec:gw_model}) and a Shapelet model (\autoref{subsec:glitch_model}) for glitches. Second, rather than using standard Time Series to Image (TTI) conversions as \cite{lisa_cnn_2024} does, we utilize time-frequency representations (\autoref{subsec:q-transform}) of the data for search and classification. Finally, whereas \cite{lisa_cnn_2025} uses an encoder-decoder based framework for source separation of overlapping GWs and glitches, our work explores the efficacy of CNN models for accurate identification of such cases. 

Our pipeline could support a range of downstream analyses of LISA data. Rapid classification of detected transients could help with model selection for the global fit. A low-latency implementation could enable early characterization of sources and provide useful information for time-sensitive studies. In the future, we plan to extend this framework to recognize broader LISA populations, such as EMRIs.

\section{Simulating LISA data}\label{sec:dataset_generation}

\begin{table*}
    \centering
    \begin{tabular}{|c|c|c|}\hline
    \rowcolor{LightCyan}
         \textbf{Parameter}&  \textbf{Units}& \textbf{Sampling distribution}\\\hline
 \rowcolor{LightCyan}
 \multicolumn{3}{|c|}{\textbf{Gravitational Wave Parameters}}\\\hline
         Chirp mass of two black holes $\mu$&  Solar masses& $\log \mu \sim \mathcal{U}(4, 7)$\\\hline
         Mass ratio of two black holes $(q)$&  Unitless& $q\sim \mathcal{U}(1, 5)$\\\hline
         Luminosity distance $(d)$&  Megaparsecs (Mpc)& $\log d \sim \mathcal{U}(3, 5)$\\\hline
         Spins of two black holes $(\chi_1, \chi_2)$&  Ratio of maximum spin& $\chi_1, \chi_2 \sim \mathcal{U}(-1, 1)$\\\hline
         Inclination $(\iota)$&  Radians& $\cos \iota \sim \mathcal{U}(-1,1)$\\\hline
         Source ecliptic latitude $(\beta_{\text{GW}})$ and longitude $(\lambda_{\text{GW}})$&  Radians& $\sin\beta_{\text{GW}} \sim \mathcal{U}(-1, 1)$ and $\lambda_{\text{GW}} \sim \mathcal{U}(0, 2\pi)$\\ \hline
         \rowcolor{LightCyan}
 \multicolumn{3}{|c|}{\textbf{Glitch Parameters}}\\\hline
 Amplitude $(A)$& Acceleration $(\nicefrac{\text{m}}{\text{s}^2})$&$\log|A| \sim \mathcal{U}(-14, -11)$\\\hline
 Timescale $(\beta)$& Seconds&$\log\beta \sim \mathcal{U}(0, 4)$\\ \hline
 Injection point& Not applicable&Randomly chosen from $\text{TM}_{ij}$ for $i,j\in \{1,2,3\}$\\\hline
    \end{tabular}
    \caption{Parameters that describe the GW and glitch models we used in this study. A set $\{\mu, q, d, \chi_1, \chi_2, \iota, \beta_{\text{GW}}, \lambda_{\text{GW}}\}$ completely defines a GW waveform (\autoref{subsec:gw_model}) and a set $\{A, \beta, \text{inj-point}\}$ completely defines a glitch (\autoref{subsec:glitch_model}). Values of these parameters were sampled from their respective sampling distributions to produce training and testing datasets (see \autoref{subsubsec:search_filtering}).}
    \label{tab:gw_glitch_params}
\end{table*}

The final operational pipeline will detect and categorize transients within a single time series containing several days to months of data. Since real LISA data are not yet available, we constructed the training dataset from simulations produced with the \codeword{lisainstrument} Python package \citep{lisainstrument}. Instead of creating one long simulation covering multiple months, we train our classifier on many shorter, independent simulations, each containing at least one distinct injected transient. Generating this dataset involved the following steps:

\begin{enumerate}
    \item Generate several simulations of LISA TDI channels, each with one or more GW and/or glitch injected.
    \item Create Q-transforms of data segments containing the transients of interest.
    \item Validate whether the search algorithm can detect the transients present in each Q-transform. Keep only the cases where all injected transients could be recovered.
\end{enumerate}

This last step is essential as the classifier will only receive Q-transforms where the search detected a potential event. Therefore cases where the search cannot recover the event are not relevant to the classifier. The only exception to this would be simulations where no transients were injected (see \autoref{subsec:noise_simulations}). These examples were included to prevent the classifier from learning structures in background noise as transient features. 

The \codeword{lisainstrument} package \citep{lisainstrument} simulates the noises of the individual subsystems and combines them to produce the realistic approximation of the LISA raw interferometer measurements which is then processed via the \codeword{pytdi} package \citep{pytdi} to produce the aforementioned TDI channels. To construct a dataset representative of realistic LISA observations, we generate readouts based on a time-evolving spacecraft configuration governed by realistic orbital dynamics. The orbit of the LISA constellation over a four year mission period is generated using the \codeword{lisaorbits} package \citep{lisaorbits}.

For this study we simulated around 6000, twelve hour segments of TDI readout from the $A,E$ channels, with a sampling rate of $1$ second. Each simulation includes either only a GW, only a glitch, both a GW and a glitch, or no transient at all. For simulations that included a single transient event, the injection was performed at the 6-hour mark. \autoref{subsec:double_transients} describes the injection procedure for simulations with multiple transients. The starting time of each simulation was chosen at random from the four-year time range over which the orbits were calculated. Since the actual mission will have transients occurring under different spacecraft configurations, resulting in different signal behavior in TDI channels, this approach is used to make our classifier robust to such changes. Each transient type considered in this study is described by a parametric model with one or more free parameters. We sample several values of these parameters from their respective distributions, each sample defining the injected transient in one simulation. The following subsections will outline the models used to describe our injected transients and \autoref{subsubsec:search_filtering} will detail our procedure to sample parameter values from their respective assumed priors shown in \autoref{tab:gw_glitch_params}.

\subsection{Gravitational wave simulations}\label{subsec:gw_model}
The GWs waveforms in this study are generated using the \codeword{pycbc} package \citep{pycbc} using the \codeword{IMRPhenomXHM} approximant \citep{imrphenomxhm}. The GW waveforms of merging MBHBs are described by 7 independent parameters as shown in \autoref{tab:gw_glitch_params}. The chirp mass is defined as
\begin{equation}
    \mu = \dfrac{(m_1m_2)^{\frac{3}{5}}}{(m_1+m_2)^{\frac{1}{5}}},
    \label{eqn:chirp_mass}
\end{equation}
where $m_1,m_2$ are the masses of the two BHs. Furthermore, the \codeword{IMRPhenomXHM} approximant generates waveforms for MBHBs with spins aligned with their orbital angular momentum and hence the sampled spins are chosen accordingly. Each injected GW waveform defined by the set $\{\mu, q, d, \chi_1, \chi_2, \iota, \beta_{\text{GW}}, \lambda_{\text{GW}}\}$. It should be noted that the mass of the two black holes are computed using the sampled values of $\mu, q$ and that the polarisation of the MBHB system is set to the default value of zero in each injection. 

\newpage
\subsection{Glitch simulations}\label{subsec:glitch_model}
The LPF mission, described in \autoref{subsec:lisa_pathfinder}, identified and characterized instrumental glitches now expected in the full scale LISA mission \citep{lpf_glitches}. These transients are currently best modeled with Shapelet functions \citep{lpf_glitches} which is expressed as
\begin{equation}
    \psi_n(t) = \dfrac{2t}{n}e^{-\frac{t}{n}}L^1_{n-1}\Big(\dfrac{2t}{n}\Big)\Theta(t),
    \label{eqn:shapelet}
\end{equation}
where $L^1_{n-1}$ denotes the generalized Laguerre polynomial and $\Theta(t)$ is the Heaviside step function. The LPF glitches were unexpected deviations in the acceleration of the test-masses as a function of time. These were modeled using shapelets of \autoref{eqn:shapelet} with $n=1$, scaled by a factor $A$ as
\begin{equation}
    g(t) = 2At'e^{-t'}\Theta(t'),
    \label{eqn:glitch_acc}
\end{equation}
where $t'=\nicefrac{(t-t_{\text{inj}})}{\beta}$, $g(t)$ is the test mass acceleration, $A$ is also referred to as the \textit{amplitude}, $\beta$ is the glitch timescale and $t_{\text{inj}}$ is the injection time. The beat note fluctuations induced in the interferometers due to $g(t)$ can be obtained by integrating \autoref{eqn:glitch_acc} as \citep{lpf_glitches}
\begin{equation}
    \dfrac{\Delta \nu (t)}{\nu_0} = \dfrac{2A\beta}{c}\Big(1-e^{-\frac{t-t_{\text{inj}}}{\beta}}\Big(\dfrac{t-t_{\text{inj}}}{\beta}-1\Big)\Big)\Theta\Big(\dfrac{t-t_{\text{inj}}}{\beta}\Big),
    \label{eqn:glitch_freq}
\end{equation}
where $\nu_0=2.816\times 10^{14}$ Hz is the nominal laser frequency and $c$ is the speed of light. \autoref{eqn:glitch_freq} represents the model used to inject glitches at one of the 6 test mass interferometers at random in our simulations. Each injected glitch is defined by the set $\{A, \beta, \text{inj-point}\}$.

\subsection{Double transient simulations}\label{subsec:double_transients}

Since instrumental glitches maybe present in the Q-transform frame during the passing of a gravitational wave, it is important to consider cases involving the presence of both these transients. To simulate such cases, we define a separation time $t_{\text{sep}}$ which is proportional to the elapsed time between the GW merger and the peak of the glitch. For a given $t_{\text{sep}}$, the GW is injected at $6 + t_{\text{sep}}$ hours and the glitch at $6 - t_{\text{sep}}$ hours. The values of $t_{\text{sep}}$ were sampled uniformly at random from $-3$ to $3$ hours with negative values including cases where the GW occurs before the glitch. Hence our dataset includes cases where a glitch occurs within 6 hours of a GW merger. 

\subsection{Gaussian noise simulations}\label{subsec:noise_simulations}
To prevent our classifier from learning structures in the background noise as evidence for transients, we included simulations without any injected transients in our dataset. These train the classifier to differentiate between cases involving the absence and presence of transients. The background noise in every simulation follows the expected LISA noise curve \citep{noise_asd, lisa}.

\begin{figure*}
    \centering
    \includegraphics[width=\textwidth]{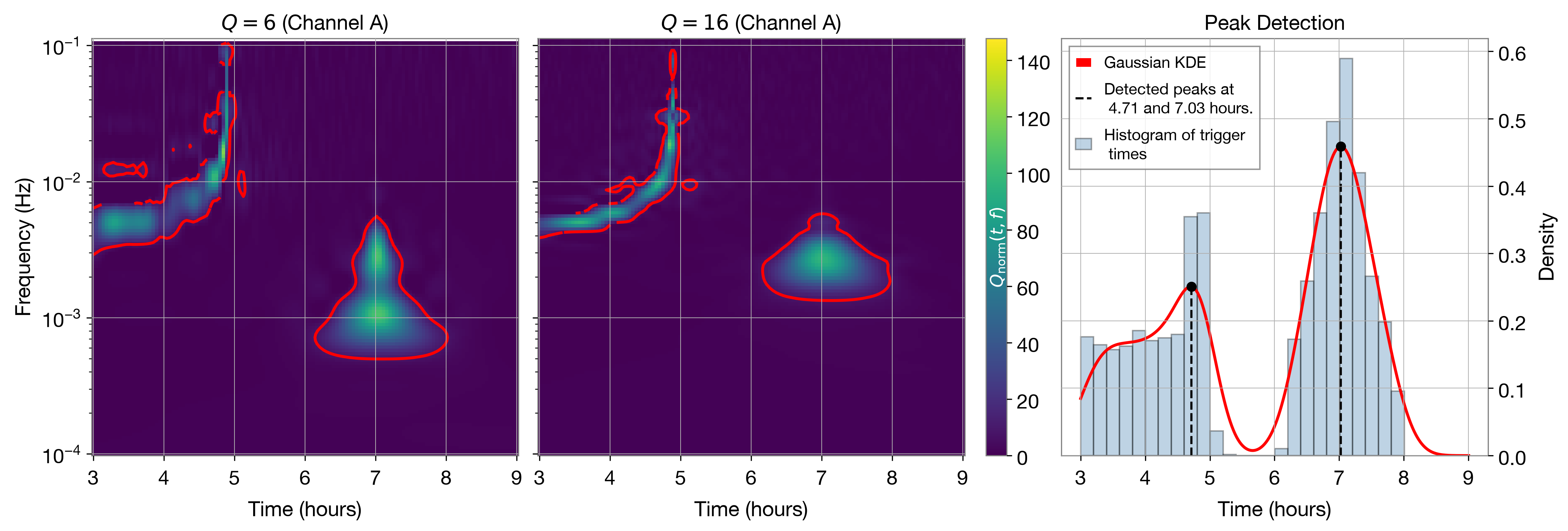}
    \caption{The two panels on the left show Q-transform representations with different $Q$ values of the same GW and glitch simulation (see \autoref{sec:dataset_generation}). It can be seen that a lower $Q$ value, better represents shorter duration transients (glitch at $t=7$ hours in this case). On the other hand, longer duration transients (GW merging close to $t=5$ hours) are better represented by higher $Q$ values. The red contours divide areas where the energy (\autoref{eqn:energy}) is greater than 8 and less than 8. The rightmost panel displays a histogram of the time distribution for all triggers in both images overlaid with the Gaussian KDE. Note that the histogram had no role in computing the KDE and is simply shown for comparison. Peaks with a prominence exceeding 0.05 in density are identified as transients, and the times at which they occur are recorded as transients.}
    \label{fig:search_algorithm}
\end{figure*}

\section{The Search Algorithm}\label{sec:search_algorithm}

As mentioned earlier, our pipeline searches for transients in Q-transforms of 6 hour segments of data in LISA TDI channels $A,E$. \autoref{subsec:q-transform}, outlines our implementation of the Q-transform followed by a description of the search algorithm in \autoref{subsec:search_algorithm}.

\subsection{Q-transforms}\label{subsec:q-transform}
Similar to spectrograms, Q-transforms plot the evolution of the frequency content in a signal as a function of time. However, contrary to a conventional spectrogram, Q-transforms are computed over a non-uniform time frequency grid, controlled by a parameter $Q$ \citep{q_transform}. Prior to applying a Q-transform, we whiten our timeseries data with the one sided power spectral density (PSD) of the appropriate TDI channel which is estimated using the Welch method with the Hann window \citep{welch}.

To obtain the Q-transform, we start with a frequency range $[f_{\text{min}}, f_{\text{max}}]$ and a time range $[t_{\text{min}}, t_{\text{max}}]$ to analyze. Starting with $f_{\text{min}}$, we iteratively choose the frequency grid points up to $f_{\text{max}}$ using
\begin{equation}
    f_{i+1}=f_i + 2\sqrt{2}\dfrac{f_i}{Q}.
\end{equation}
For each $f_i$, a time grid in the chosen time range is computed using
\begin{equation}
    t_{j+1}(f_i)=t_j(f_i) + \dfrac{1}{\sqrt{2}\pi}\dfrac{Q}{f_i}.
\end{equation}
This creates a time-frequency grid where the frequency bin width ($\Delta f_i$) and time bin width ($\Delta t_i$) at a given $f_i$ are inversely and directly proportional to $\nicefrac{Q}{f_i}$ respectively. For a whitened signal $x(t)$, the Q-transform power over the grid is hence defined to be
\begin{equation}
    Q(t_i, f_i) = \dfrac{1}{N}\Bigg|\sum_{t=t_{\text{min}}}^{t_{\text{max}}}x(t)\cdot w(t, t_i, f_i)\Bigg|^2.
\end{equation}
Here $N=\sum_{t=-\infty}^{\infty}w(t,t_i,f_i)$ is the normalization constant and $w(t,t_i,f_i)$ is known as the window function which is given by
\begin{equation}
    w(t, t_i, f_i) =
    \begin{cases}
        e^{\frac{-(t-t_i)^2}{2\tau^2}}e^{-2\pi i f_i (t-t_i)}\ \text{for}\ t-t_i\in [- 4\tau, 4\tau]\\
        0\ \text{otherwise},
    \end{cases}
    \label{eqn:window}
\end{equation}
\noindent where $\tau = \frac{Q}{\sqrt{2}\pi f_i}$. The result is also normalized along the time axis as
\begin{equation}
    Q_{\text{norm}}(t,f)=\dfrac{Q(t,f)}{\langle Q(t,f)\rangle_{f}},
\end{equation}
where $\langle \rangle_f$ denotes the median over constant $f$. Once $Q_{\text{norm}}(t, f)$ is computed for a given simulated signal, it is interpolated on a uniform time-frequency grid with the chosen bounds $[f_{\text{min}}, f_{\text{max}}], [t_{\text{min}}, t_{\text{max}}]$ and resolution $128\times 128$. An example is shown in \autoref{fig:search_algorithm} which shows the Q-transform of a window containing a GW and a glitch for two $Q$ values: 6 (left) and 16 (right). Our implementation of the Q-transform differs from a similar approach by \cite{alan_search} in the choice of window function. Whereas they used a bi-square window, we use a sinusoidal wavelet given by \autoref{eqn:window}.

Q-transforms are advantageous over conventional spectrograms while dealing with frequency ranges spanning several orders of magnitude as in our case. This is because lower frequency components are captured over a larger period of time as compared to larger frequency components. A non uniform grid accounts for this by allowing the window function to operate on a longer part of the signal at lower frequencies.

\begin{figure*}
    \centering
    \includegraphics[width=0.9\textwidth]{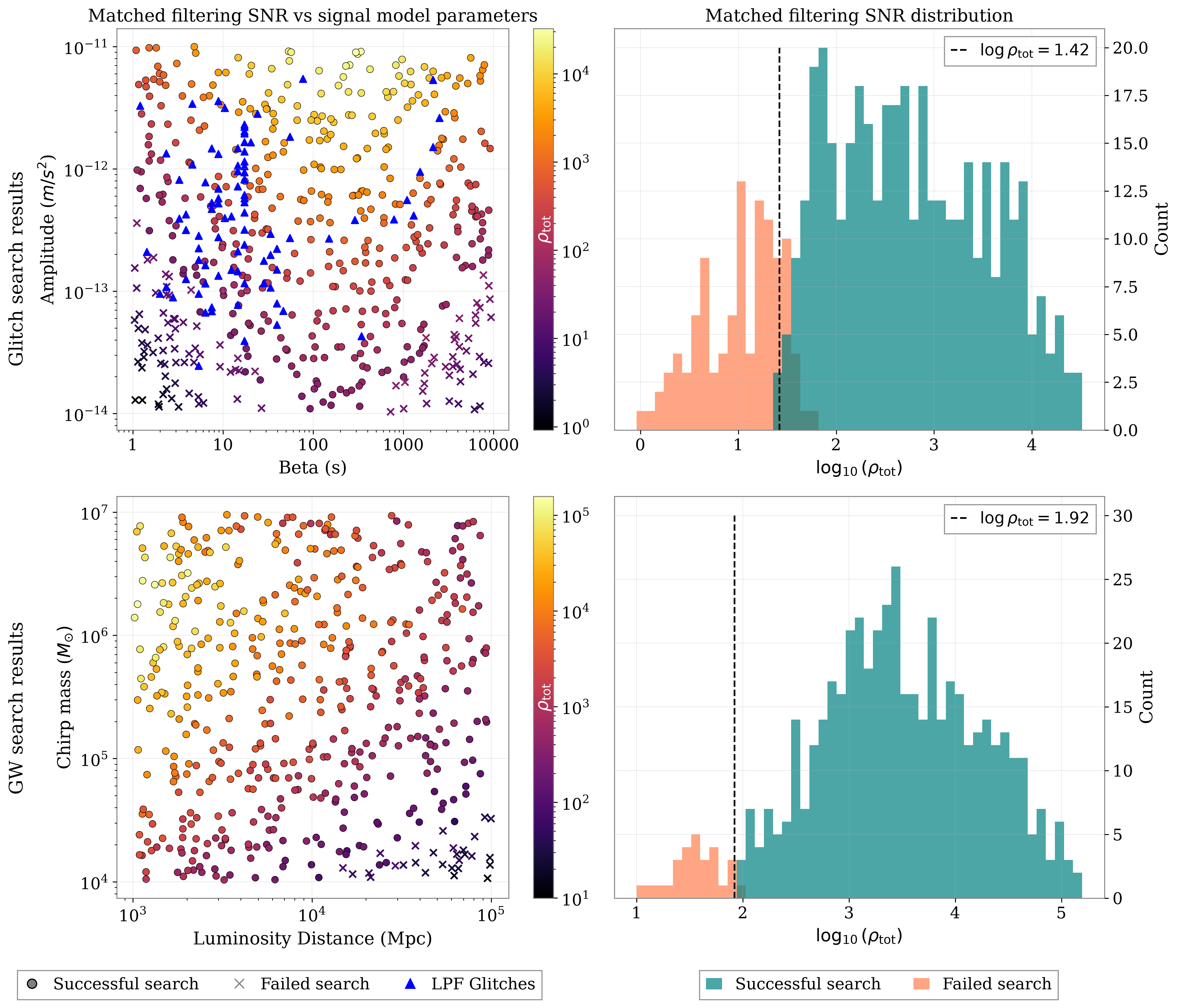}
    \caption{
    \textit{Top left:} Sampled glitch model parameter values from their respective distributions in \autoref{tab:gw_glitch_params}. \textit{Bottom left:} Sampled GW chirp mass and luminosity distance values from their respective distributions in \autoref{tab:gw_glitch_params}. In either panel, the blue circles represent the simulations where the transient was recovered by our search algorithm and the orange crosses denote the cases where it was not. In the left panel, the blue triangles represent the fit results of the LPF glitches \citep{lpf_glitches}. It can be seen that our search algorithm can successfully recover most of them. \textit{Top right:} Two histograms showing the distribution of matched filter SNRs of the injected glitches. Each histogram is colored based on the ability of the search to recover the injected signals lying within the SNR bin.
    \textit{Bottom right:} Same as top right but with the distribution of matched filter SNRs of injected GWs of merging MBHBs.}
    \label{fig:filtering}
\end{figure*}

\subsection{Excess energy based search}\label{subsec:search_algorithm}
Q-transforms taken with a single $Q$ value cannot represent the vastly different GW and glitch morphologies expected in LISA. Therefore it is usually helpful to take multiple Q-transforms of the same data with different values of $Q$. This can be seen in \autoref{fig:search_algorithm} where $Q=6$ represents a short duration glitch, better than $Q=16$. The opposite is true for the longer duration GW which is also present. Empirically, we determined that $Q=6,16$ are sufficient to represent expected transient morphologies over the LISA frequency range. Hence we compute a three-dimensional grid $X(t, f, Q)$ of every 6 hour segment of channels $A,E$ using $Q=6,16$. This results in a grid with shape $128\times 128\times 4$ where the first two images represent the $Q=6,16$ representation of the $A$ channel respectively and the last two represent the same for the $E$ channel. The \textit{`energy'} on this grid is defined as
\begin{equation}
    \rho(t,f,Q) = \sqrt{X^2(t,f,Q)-1},
    \label{eqn:energy}
\end{equation}
where we follow a similar approach to \cite{alan_search}, and define any point with $\rho(t,f,Q)\geq 8$ as \textit{triggers}. For a given collection of triggers, we record their time coordinates. We then use the \codeword{scipy} package \citep{scipy} to compute a Gaussian Kernel Density Estimate (KDE) over these trigger times ($T_i$) as given by
\begin{equation}
    \hat{f}(t)=\dfrac{1}{nh\sqrt{2\pi}}\sum_{i=1}^n e^{-\frac{(t-T_i)^2}{2h^2}},
\end{equation}
\noindent where $n$ is the number of data points and $h=\nicefrac{\sigma}{n^{\frac{1}{5}}}$, is the bandwidth chosen by Scott's rule \citep{scott_rule} for sample standard deviation $\sigma$. From the resulting KDE, we extract the local maxima with prominence exceeding $0.05$, using \codeword{find_peaks} function from the \codeword{scipy} package. The time associated with the peak is taken as the time of transient events. \autoref{fig:search_algorithm} shows an example of this process applied on a Q-transform of a simulated glitch and GW (for more details on the simulations, see \autoref{sec:dataset_generation}. The Q-transforms containing at least one transient are then passed on to the classifier for transient identification.

\section{Classifier Training}\label{sec:classifier_training}

\subsection{Feature set generation}\label{subsec:feature_set}

The classifier was trained on Q-transforms of simulated LISA transients. The following subsections describe the process by which this was achieved.

\newpage
\subsubsection{Search based filtering}\label{subsubsec:search_filtering}
As mentioned earlier in the opening of \autoref{sec:dataset_generation}, the search algorithm was used to keep only those images in the training data, where the search could recover the injected transients. To generate the dataset, first, 500 values of each GW and glitch model parameters were sampled from their respective prior distributions in \autoref{tab:gw_glitch_params}. Then we produced 12 hour simulations in which we injected either only GWs or only glitches at the 6 hour mark, using the sampled parameters. Furthermore, for each of the 500 GWs and glitches, we also generated noise-free detector responses with the signals injected at the same 6 hour mark. This was done to obtain templates for the expected signals in the LISA TDI channels which were then used to calculate the matched-filter SNR of each signal in its noisy simulation, given by \citep{match_filter}
\begin{equation}
    \rho_{\text{mf}}=4\int_0^{\infty}\dfrac{|\tilde{h}(f)|^2}{S_n(f)}df,
    \label{eqn:match_filter}
\end{equation}

\begin{figure}
    \centering
    \includegraphics[width=\columnwidth]{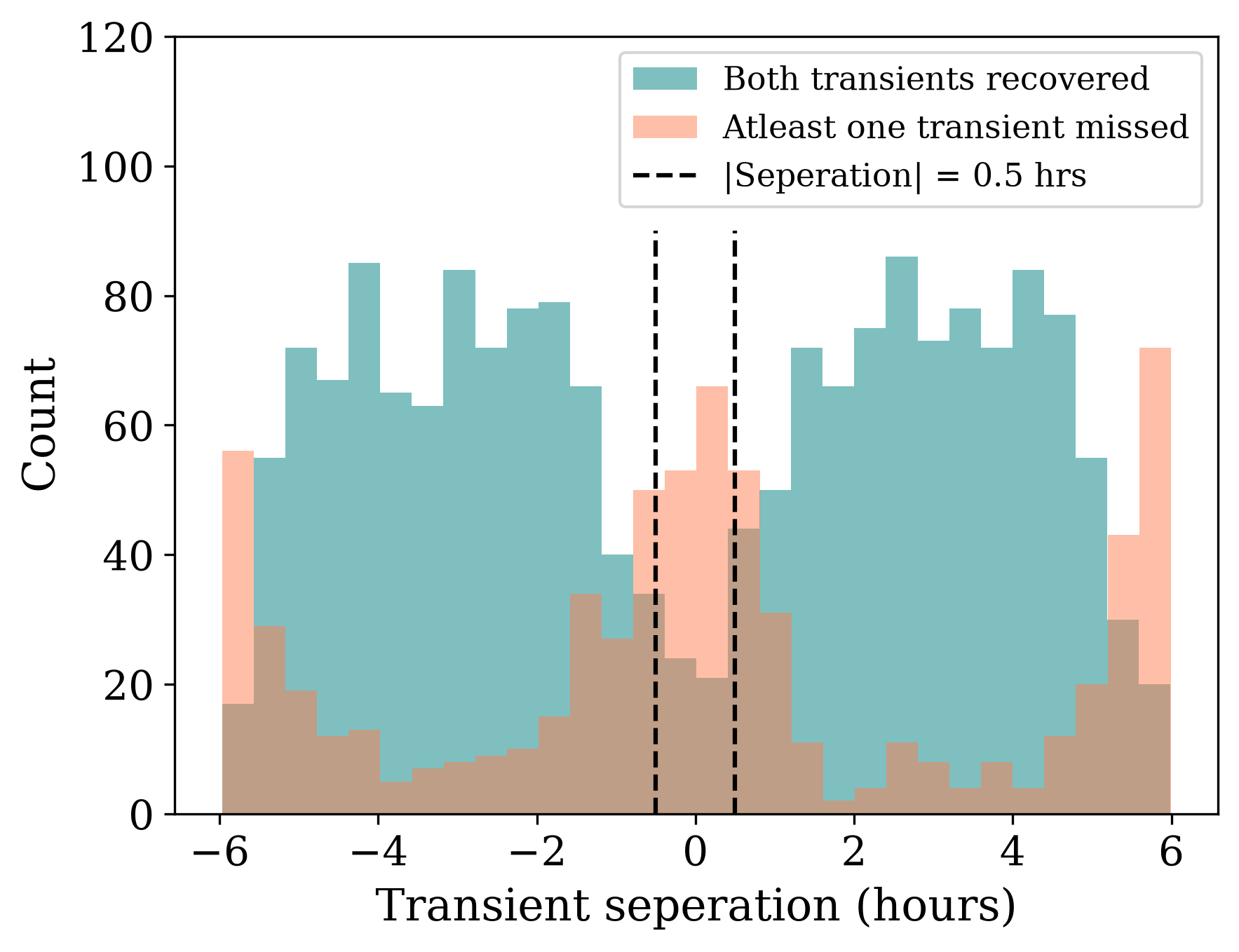}
    \caption{A histogram of separation times (in hours) between injected transients for simulations in which both transients were recovered and those in which at least one transient was missed. A negative separation denotes cases where the glitch was injected before the GW. The black dashed lines denote an absolute separation of 0.5 hours.}
    \label{fig:mixed_filtering}
\end{figure}

\noindent where $\tilde{h}(f)$ is the Fourier transform of the signal template in the LISA TDI channel and $S_n(f)$ is the one sided PSD of said channel. Although \autoref{eqn:match_filter} was originally defined by \cite{match_filter} for GW strain in LIGO data, we extend this definition for the LISA TDI channels. Since the noise in the $A,E$ TDI channels can be approximated to be uncorrelated \citep{aet_tdi}, the total match filter SNR of a signal across these channels can be defined as
\begin{equation}
\rho_{\text{tot}}=\sqrt{\rho_A^2+\rho_E^2},
\end{equation}
where $\rho_A,\rho_E$ are the signal's match filtered SNR in the $A,E$ channels respectively. The scatter plots in \autoref{fig:filtering} are color coded based on the matched filter SNR of the corresponding signals in their noisy simulations.

\begin{figure*}
    \centering
    \includegraphics[width=0.75\textwidth]{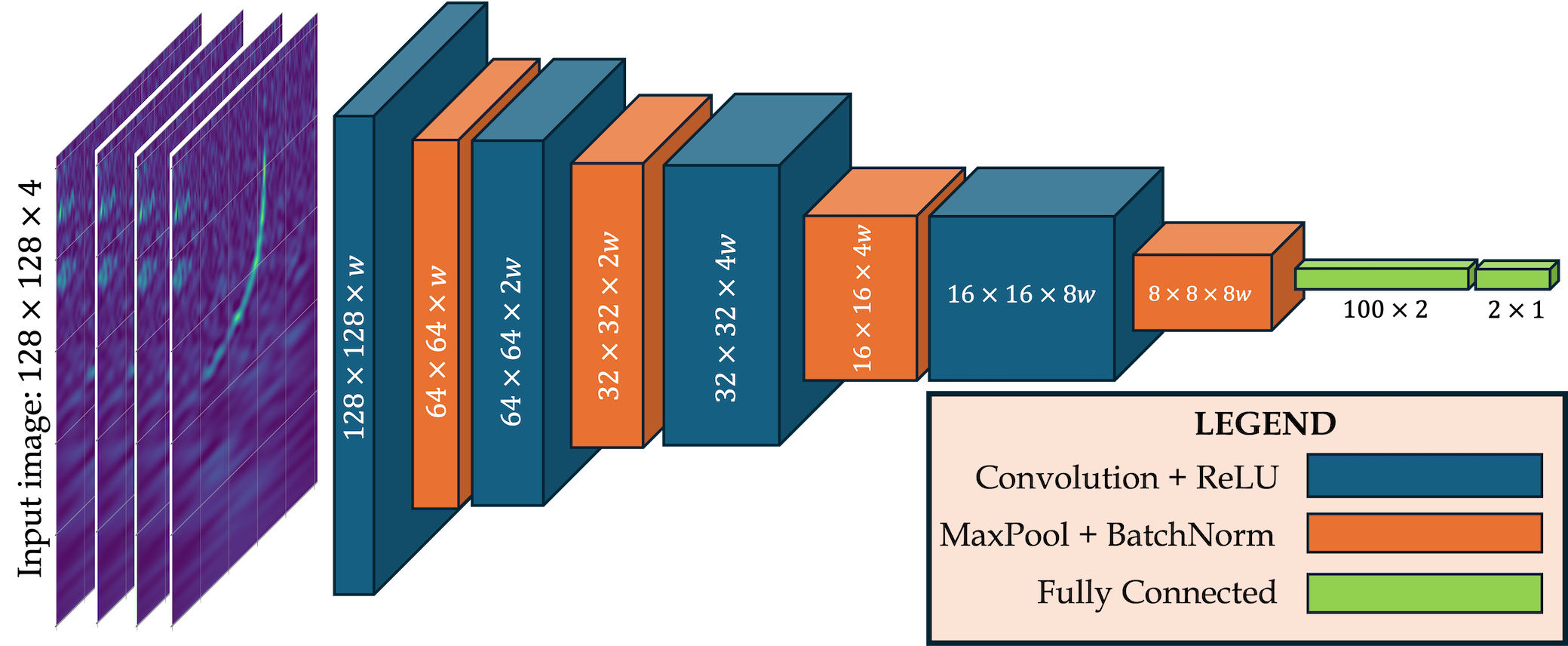}
    \caption{A pictorial representation of the CNN used in our study. The classifier takes as input, images with a shape of $128\times 128\times 4$ and outputs a $2\times 1$ column vector containing the scores for each transient being present in the image. For a given transient, a score of 1 indicates presence and a score of 0 indicates absence. The shapes of the output of each layer is also shown where $w$ is a variable hyperparameter. It must be noted that the boxes used to represent the layers are not shown to scale with respect to their labeled shapes.}
    \label{fig:model_architecture}
\end{figure*}

We applied the search algorithm to the Q-transforms of each of the 500 simulated GWs and glitches, allowing us to divide the parameter space in \autoref{fig:filtering}, into regions were the transient was detectable and where it was not. The distribution of the matched filter SNRs of the simulated signals is also shown in the right column of \autoref{fig:filtering} and is color coded based on whether the search could recover the signal. We see that for glitches, the matched filter SNR at the boundary of detectability is approximately $26.3$ and that for GWs is around $83.2$. This quantifies the effectiveness of the search algorithm presented in this paper. The left column of \autoref{fig:filtering} also shows the parameters of the signals that can be recovered by our algorithm. In the plot of the glitch parameter space in \autoref{fig:filtering}, we also mark the fit results of the LPF glitches when modeled with Shapelets \citep{lpf_glitches}. We see that the detectable region of our search algorithm covers almost all of these points. A \codeword{RandomForest} model \citep{random_forest} was trained to learn the boundaries of the search recoverable region for both signal types and used to generate an additional 1000 GW and glitch model parameter samples from their corresponding recoverable regions (circular points in \autoref{fig:filtering}). This resulted in 1,474 glitch simulations and 1,491 GW simulations to be considered for our dataset.

For double transient simulations, around 1500 GW and glitch model parameters were sampled from the recoverable regions defined by the aforementioned \codeword{RandomForest}. Furthermore, in this case, only the simulations where the search algorithm could recover both signals were included in the training dataset. This was done to ensure that the classifier’s misclassifications could be attributed to indistinguishable signal features rather than to either signal being undetectable by the search algorithm. Instances in which the search failed to recover both transients occurred either because the transients overlapped too closely in time or because one transient was significantly louder than the other. This is illustrated in \autoref{fig:mixed_filtering}, which shows a histogram of separation times between injected transients for simulations in which both transients were recovered and those in which at least one transient was missed. It can be seen that the search is inefficient at recovering transients which are separated by less than 0.5 hours. This can be interpreted as the resolving power of the search algorithm. Future work would attempt to improve this by using shorter duration Q-transforms with finer resolution. Furthermore, the search also fails to recover transients with separations close to 6 hours. This is expected as, 6 hours is the time-width of every Q-transform and transients separated by this amount lie at the boundaries of the image and can be missed. In an operational pipeline, this will not effect the performance as the same transient would be fully visible in an adjacent Q-transform. Post filtering, we had 1,346 double transient simulations to be used for generating Q-transforms.

\subsubsection{Transient centering}\label{subsubsec:transient_centering}

The position of the transient within a Q-transform is arbitrary. A good classifier must be able to identify a transient regardless of where it is in the image. Therefore, for every transient, the Q-transform window length was chosen to be 6 hours. This provides freedom to translate the window in the 12 hour long simulation, allowing for various relative transient placements within the image. To train the classifier on a wide range of transient centerings, we sample five values of the centering time for each image. The sampling was done in different ways for single and double transient cases.

In case of simulations with only one present transient, five values of the Q-transform centering times were chosen at random between the $3$ and $9$ hour mark. Since the transient itself was always injected at the 6 hour mark, this results in different positioning of the signal in each Q-transform. Doing the same for the multiple transient case involves a slight modification as some centering choices can push one of the transients off the image which impact classification. To prevent this, the five centering times were sampled randomly between the $3+t_{\text{sep}}$ and $9-t_{\text{sep}}$ where $t_{\text{sep}}$ is the seperation parameter introduced in \autoref{subsec:double_transients}. This process allows for the classifier to see multiple examples of the same transient centered at a different points during training, ensuring robustness to offsets in transient placement within an image. Hence we generate five Q-transforms per simulated transient. Our final training dataset consisted of 7,500 images with Gaussian noise, 7,370 images with glitches and noise, 7,455 images with GWs and noise and 6,730 images with both transients in Gaussian noise. This resulted in a total dataset size of 29,055 images.

\begin{figure*}
    \centering
    \includegraphics[width=\textwidth]{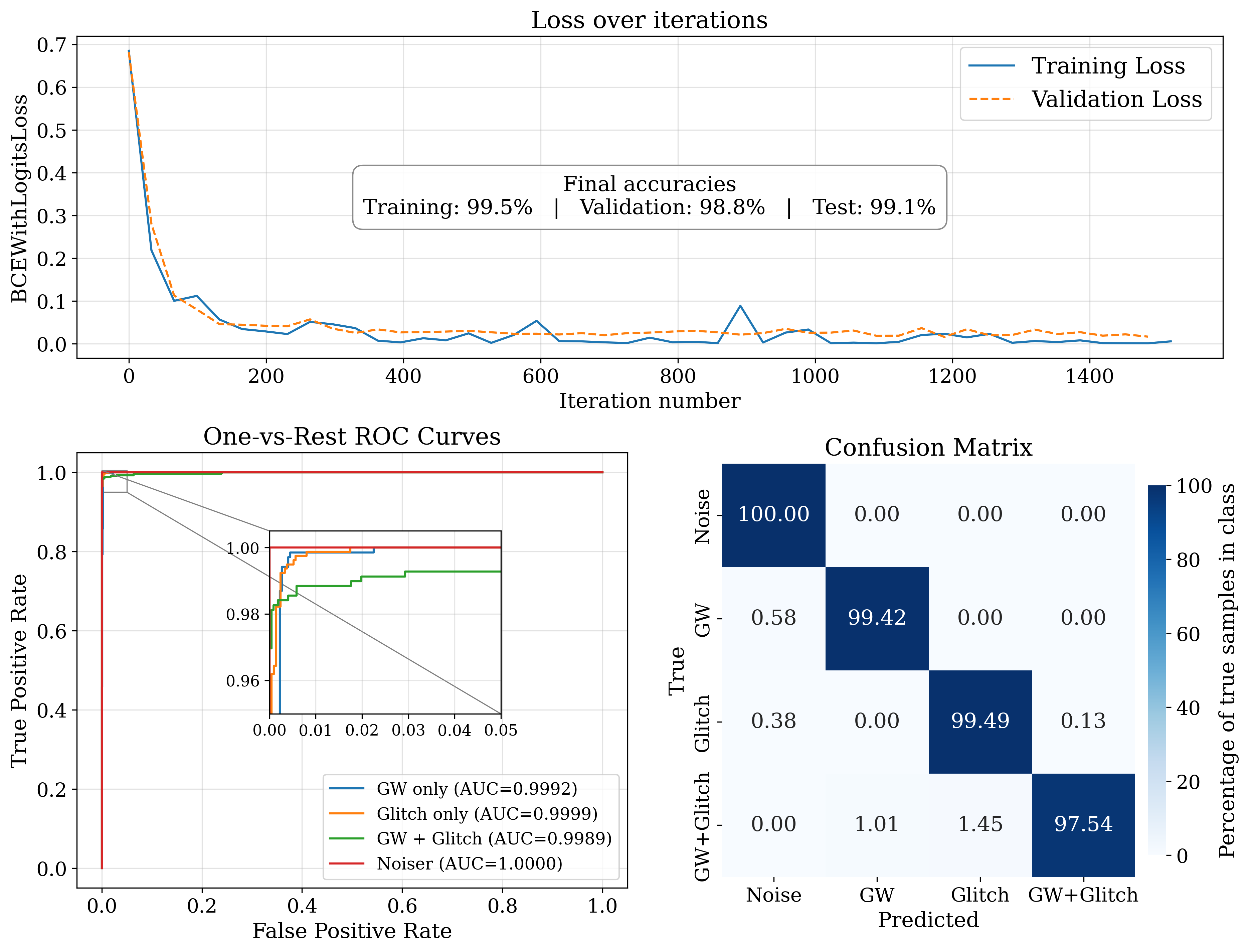}
    \caption{The first row shows the training progress in terms of the evolution of the training and validation loss over iterations. The left plot in the second row shows the post-training ROC of the classifier on the testing set. Since a ROC curve is defined for binary classification, we plot four one-vs-rest ROC curves, each showing how well the classifier distinguishes one of the four possible cases from all the remaining cases. The right plot in the second row shows the confusion matrix on the testing set after training was complete. Each entry shows the percentage of samples from a given true class that are classified into each predicted class.}
    \label{fig:model_performance}
\end{figure*}

\subsection{Classifier Architecture}\label{subsec:architecture}
Our CNN classifier consists of four convolutional layers, four pooling layers and two fully connected layers, as shown in \autoref{fig:model_architecture}. It takes as input, a $128\times 128\times 4$ Q-transform grid described in \autoref{subsec:search_algorithm}. Every image is normalized using
\begin{equation}
    \vec{x}_{\text{norm}} = \dfrac{\vec{x}-\mu}{\sigma},
\end{equation}
where $\vec{x}$ is the original grid and $\mu, \sigma$ are the mean and standard deviation, respectively across the entire grid. This ensures that every image the classifier sees has zero mean and unit standard deviation which prevents it from having to adjust weights to account for the varying loudness of different transients. However, this preserves the relative brightness of the same transient in different TDI channels and $Q$ values used for the transform.

The width parameter $w$ that controls the layer shapes is a variable hyperparameter. Every convolutional layer is followed by a \codeword{ReLU} activation. The output is then fed to a pooling layer followed by batch normalization. The first convolutional layer increases the number of channels from $4$ to $w$ and every following convolutional layer doubles this value. A kernel size of $3\times 3$ was used along with a padding of $1$. The pooling layers use the \codeword{MaxPool2D} algorithm for every $2\times 2$ block of pixels which effectively reduces the input size by half. A batch normalization is also applied after every pooling layer. The output of the final pooling layer has a shape of $8\times 8\times 8w$ which is recast into a column vector of shape $512w\times 1$ and passed through two fully connected layers of shapes $512w\times 100$ and $100\times 2$. The output is a $2\times 1$ column vector, to which, a sigmoid activation function is applied. This is given by
\begin{equation}
    p_i= \sigma(y_i)= \dfrac{1}{1+e^{-y_i}},
    \label{eqn:sigmoid}
\end{equation}
where $i\in\{1,2\}$ and $y_i$ are the elements of the output vector and $p_i$ is the sigmoid output. This sigmoid output is converted to a transient score (as described in \autoref{fig:pipeline} and \autoref{fig:model_architecture}) if it crosses a certain threshold. In this study if $p_i\geq 0.5$, the score equals 1 and 0 otherwise. A CNN architecture similar to the one described here was also used by the authors of the GravitySpy model \citep{gravityspy}.

\newpage
\section{Classifier performance}\label{sec:classifier_performance}

\subsection{Training results}
For the purposes of classifier training, our dataset was split into training, validation and testing sets with $70\%, 20\%$ and $10\%$ of the generated images belonging to each set respectively. The split was achieved such that all time offset images of the same transient belonged to the same set. This prevents the classifier from seeing the same transients while training and testing which could falsely inflate the accuracy metrics. The training progress was tracked using the Binary Cross Entropy with Logits (\codeword{BCEwithLogits}) loss function which can be expressed for a two class system as
\begin{equation}
    l(\vec{y}, \vec{t})= -\dfrac{1}{2}\sum_{i=1}^2[t_i\ln\sigma(y_i)+(1-t_i)\ln(1-\sigma(y_i))],
\end{equation}
\noindent where $\vec{y}, \vec{t}$ are the raw output (logit) and target vectors with $y_i, t_i$ representing their components \citep{bcewithlogits}. The loss was minimized using the \codeword{AdamW} optimization algorithm \citep{adamw}. The \codeword{pytorch} package \citep{pytorch} was used to train several classifiers with various values of hyperparameters such as model width, learning rate, number of training epochs and batch size. We observed that a model width $w=6$, trained on $5$ epochs with a learning rate of $0.001$ using a batch size of $64$ produced the best results. The first row of \autoref{fig:model_performance} shows the evolution of the loss on the training and validation set over iterations. The close agreement between both curves suggests that the classifier generalizes well and does not exhibit significant overfitting. The final accuracies after training are also shown in the first row of \autoref{fig:model_performance}. 

As mentioned earlier, the classification threshold used to obtain these accuracies is $0.5$. In other words, an image is said to contain a transient, if $p_i$ from \autoref{eqn:sigmoid} is greater than $0.5$. However, this threshold is arbitrary and hence, we also plot the Receiver Operating Characteristic (ROC) curve of the trained classifier on the testing set. The ROC curve is defined for a binary classification task. However, our classifier was trained to recognize four cases. Therefore we plot four ROC curves, one for each case, treating the said case as the positive class and the remaining three cases as the negative class. The results are shown in the left plot of the second row in \autoref{fig:model_performance}. The area beneath each curve (also shown in \autoref{fig:model_performance}) represents how consistently the classifier performs as the classification threshold varies, with a value of $1$ indicating ideal performance.

The high classification accuracies demonstrate that our chosen Q-transform representation contains sufficient discriminative features to differentiate between various detectable transients. The right plot of the second row in \autoref{fig:model_performance} shows the confusion matrix on the testing set after training. The value in row $i$ and column $j$ represents the fraction of images belonging to transient class $i$ that are classified as class~$j$. 

\begin{figure}
    \centering
    \includegraphics[width=0.95\columnwidth]{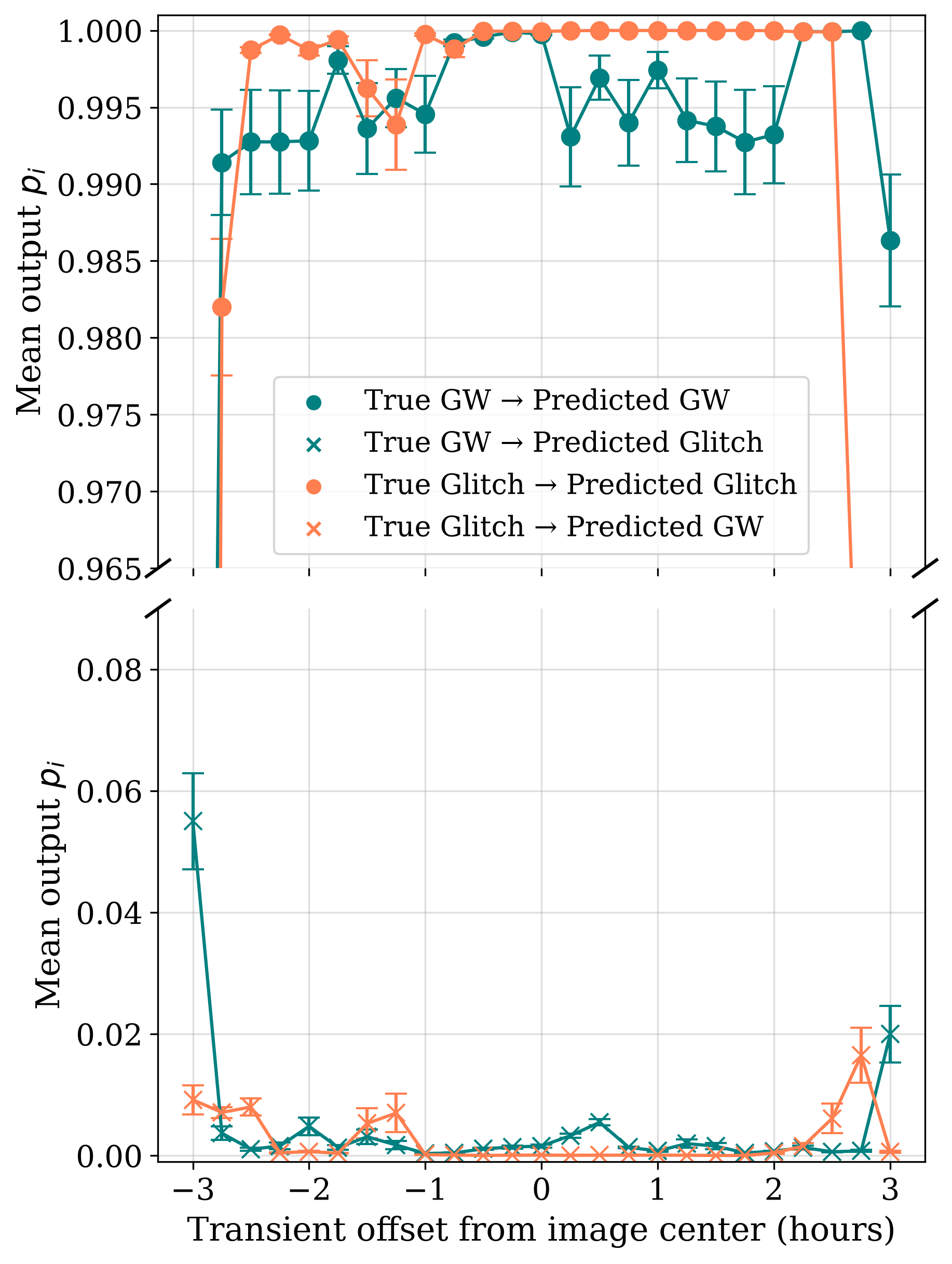}
    \caption{The $y$ axis denotes the mean output $p_i$ (\autoref{eqn:sigmoid}) on batches of images containing transients at the same offset (denoted on the $x$ axis). The error bars denote the Standard Error of the Mean (SEM).}
    \label{fig:time_centering}
\end{figure}

\subsection{Robustness to shifts in transient location}\label{subsec:time_centering}

We also test the  robustness of the classifier to shifts in the positioning of transients in the image. In an ideal classifier, less variability is favored as the detected transient position in a Q-transform could be arbitrary and should not have to be fine tuned to ensure accuracy. 

As described in \autoref{subsubsec:transient_centering}, the training dataset was designed to minimize the dependence of predictions on transient location in the image. To test the effectiveness of this approach, we focus on the correctly identified single transient images in the testing set. For every case, we produced 25 additional images, placing the transient at different offsets from the image center that ranged uniformly from $-3$ to $+3$ hours (a negative offset denotes a leftward shift). This tests the classifier's sensitivity to a resolution of around 15 minutes. The average $p_i$ (sigmoid output for each transient) on every batch of images with the same transient placement was recorded. 

The results are shown in \autoref{fig:time_centering}. Since the classifier outputs a two component vector, \autoref{fig:time_centering} contains two curves for images of each class. The error bars represent the Standard Error of the Mean (SEM) which can be written as $\nicefrac{\sigma}{\sqrt{n}}$. Here $\sigma$ is the standard deviation of the distribution of the classifier's $p_i$ output on images with the same transient offset and $n$ is the number of samples in said distribution. A higher variance in the mean prediction as a function of time signifies poorer stability. Our results in \autoref{fig:time_centering} show a fairly stable performance on the correctly classified transients, regardless of their position within the image. The performance only drops at the edge cases which is expected as centering the transient at the edge of the image might result in some loss of information. 

\section{Discussion}\label{sec:discussion}

\subsection{Conclusions}\label{subsec:conclusions}

The core findings of this paper can be divided into three main components. First, \autoref{fig:filtering} illustrates the areas of the GW and glitch parameter spaces that are, in principle, detectable by a burst search algorithm based on Q-transforms. This gives a fair idea on the types of glitches and GWs that could be quickly identified in the actual mission. \autoref{fig:filtering} also illustrates the threshold matched-filter SNR above which our algorithm can detect signals. Second, the high classification accuracies of $>98\%$  achieved by our pipeline (see \autoref{fig:model_performance}), demonstrate that among the detectable transients, the Q-transform representations contain enough morphological information to allow for classification using a four-layer CNN. Lastly, our pipeline is computationally efficient with minimal runtimes for highly accurate results. We performed our analysis on a MacBook equipped with an Apple M4 chip and 16 GB of unified memory. Classifier training was performed using the Apple Metal Performance Shaders (MPS) backend, while all runtime measurements were obtained on the same hardware. The major bottlenecks were in simulating LISA data and classifier training. Since these wont be a factor for an operational pipeline, we will not consider these in our runtime computation. Generating one 6 hour Q-transform grid described earlier required around $0.8$ seconds. For one month of TDI data from the $A,E$ channels, the search algorithm will generate 144 such transforms with an overlap of 1 hour. Therefore this is expected to take around 2 minutes. The SNR based search itself is vectorized and does not contribute significantly to the runtime. A forward pass of the classifier on a single image takes around 0.01 seconds using the MPS GPU. Assuming every image produced by the search contains a transient, this implies an upper bound on the classification runtime to be around 1.4 seconds. Hence, the pipeline can process one month of LISA data in approximately 2 minutes on a laptop, with the primary bottleneck being the computation of the Q-transforms. This high speed performance of our pipeline make it advantageous for low latency data analysis and search for the global fit.

\subsection{Limitations}
Our methodology has some inherent limitations and some temporary limitations which shall be addressed in future work. One inherent limitation is the duplication of detected signals caused due to overlapping windows while applying the Q-transform. Although the overlaps are introduced to prevent loss of information at the edges of the images, it comes with the disadvantage of the same signal being detected twice. This could reduce efficiency and a way to group duplicate detections would be needed. Furthermore, the pipeline, as presented in this paper is only trained to classify glitches and GWs from MBHB mergers. However, the burst search algorithm will flag any power excess in a Q-transform. This also includes situations where the MBHB inspiral is within the analysis band while the merger occurs outside it. This is a temporary limitation as the classifier is not trained to recognize such images. However, we aim to address this in future work as described below.

\subsection{Further avenues of research}

Overall, our pipeline could have use cases for both low-latency and downstream analysis. Classifying the detected transients reduces the parameter space for fitting GWs, thus reducing computation times for global fit analyses. Tuning the search algorithm to generate shorter duration Q-transforms could improve runtime, which could be used as a rapid alert system for low latency searches. This would also provide finer time resolution which could allow for accurate detection of transients with a separation shorter than the current limit of 0.5 hours (\autoref{fig:mixed_filtering}). Currently our pipeline's classifier recognizes GW mergers from MBHBs and glitches modeled with shapelets. In future works, the classifier will be trained to recognize other transients LISA is expected to detect (such as EMRIs) and every combination of those transients that could appear together in a Q-transform window. The classifier would also be trained to recognize GWs from their early inspiral phase which can then be modeled to estimate their time to merger allowing for prioritization of follow up electromagnetic observations.

\section*{Acknowledgments}
Aryamann's work is supported by the CSA-FAST award and the UBC-CNRS Joint PhD Program award. Jess acknowledges support from the CRC (Tier 2). Natalia and Malvina acknowledge the support of the UBC-CNRS Joint PhD Program award. We are thankful to Dr. Mervyn Chan and Dr. Raymond Ng at the University of British Columbia for their useful guidance on good practices for constructing and training machine learning models.

\newpage
\bibliography{references}

@misc{lisainstrument,
       author = {{Bayle}, Jean-Baptiste and {Hartwig}, Olaf and {Staab}, Martin},
        title = "{LISA Instrument}",
         year = 2022,
        month = aug,
          eid = {10.5281/zenodo.7071251},
          doi = {10.5281/zenodo.7071251},
      version = {1.1.1},
    publisher = {Zenodo},
       adsurl = {https://ui.adsabs.harvard.edu/abs/2022zndo...7071251B}
}

@misc{pycbc,
       author = {{The PyCBC team}},
        title = "{PyCBC: Gravitational-wave data analysis toolkit}",
 howpublished = {Astrophysics Source Code Library, record ascl:1805.030},
         year = 2018,
        month = may,
          eid = {ascl:1805.030},
archivePrefix = {ascl},
       eprint = {1805.030},
       adsurl = {https://ui.adsabs.harvard.edu/abs/2018ascl.soft05030T}
}

@ARTICLE{aet_tdi,
       author = {{Prince}, Thomas A. and {Tinto}, Massimo and {Larson}, Shane L. and {Armstrong}, J.~W.},
        title = "{LISA optimal sensitivity}",
      journal = {\prd},
         year = 2002,
        month = dec,
       volume = {66},
       number = {12},
          eid = {122002},
        pages = {122002},
          doi = {10.1103/PhysRevD.66.122002},
archivePrefix = {arXiv},
       eprint = {gr-qc/0209039},
 primaryClass = {gr-qc},
       adsurl = {https://ui.adsabs.harvard.edu/abs/2002PhRvD..66l2002P}
}

@ARTICLE{tdi,
       author = {{Tinto}, Massimo and {Dhurandhar}, Sanjeev V.},
        title = "{Time-delay interferometry}",
      journal = {Living Reviews in Relativity},
         year = 2021,
        month = dec,
       volume = {24},
       number = {1},
          eid = {1},
        pages = {1},
          doi = {10.1007/s41114-020-00029-6},
archivePrefix = {arXiv},
       eprint = {gr-qc/0409034},
 primaryClass = {gr-qc},
       adsurl = {https://ui.adsabs.harvard.edu/abs/2021LRR....24....1T}
}

@ARTICLE{lisa,
       author = {{Colpi}, Monica and {Danzmann}, Karsten and {Hewitson}, Martin and {Holley-Bockelmann}, Kelly and {Jetzer}, Philippe and {Nelemans}, Gijs and {Petiteau}, Antoine and {Shoemaker}, David and {Sopuerta}, Carlos and {Stebbins}, Robin and {Tanvir}, Nial and {Ward}, Henry and {Weber}, William Joseph and {Thorpe}, Ira and {Daurskikh}, Anna and {Deep}, Atul and {Fern{\'a}ndez N{\'u}{\~n}ez}, Ignacio and {Garc{\'\i}a Marirrodriga}, C{\'e}sar and {Gehler}, Martin and {Halain}, Jean-Philippe and {Jennrich}, Oliver and {Lammers}, Uwe and {Larra{\~n}aga}, Jonan and {Lieser}, Maike and {L{\"u}tzgendorf}, Nora and {Martens}, Waldemar and {Mondin}, Linda and {Piris Ni{\~n}o}, Ana and {Amaro-Seoane}, Pau and {Arca Sedda}, Manuel and {Auclair}, Pierre and {Babak}, Stanislav and {Baghi}, Quentin and {Baibhav}, Vishal and {Baker}, Tessa and {Bayle}, Jean-Baptiste and {Berry}, Christopher and {Berti}, Emanuele and {Boileau}, Guillaume and {Bonetti}, Matteo and {Brito}, Richard and {Buscicchio}, Riccardo and {Calcagni}, Gianluca and {Capelo}, Pedro R. and {Caprini}, Chiara and {Caputo}, Andrea and {Castelli}, Eleonora and {Chen}, Hsin-Yu and {Chen}, Xian and {Chua}, Alvin and {Davies}, Gareth and {Derdzinski}, Andrea and {Domcke}, Valerie Fiona and {Doneva}, Daniela and {Dvorkin}, Irna and {Mar{\'\i}a Ezquiaga}, Jose and {Gair}, Jonathan and {Haiman}, Zoltan and {Harry}, Ian and {Hartwig}, Olaf and {Hees}, Aurelien and {Heffernan}, Anna and {Husa}, Sascha and {Izquierdo-Villalba}, David and {Karnesis}, Nikolaos and {Klein}, Antoine and {Korol}, Valeriya and {Korsakova}, Natalia and {Kupfer}, Thomas and {Laghi}, Danny and {Lamberts}, Astrid and {Larson}, Shane and {Le Jeune}, Maude and {Lewicki}, Marek and {Littenberg}, Tyson and {Madge}, Eric and {Mangiagli}, Alberto and {Marsat}, Sylvain and {Vilchez}, Ivan Martin and {Maselli}, Andrea and {Mathews}, Josh and {van de Meent}, Maarten and {Muratore}, Martina and {Nardini}, Germano and {Pani}, Paolo and {Peloso}, Marco and {Pieroni}, Mauro and {Pound}, Adam and {Quelquejay-Leclere}, Hippolyte and {Ricciardone}, Angelo and {Rossi}, Elena Maria and {Sartirana}, Andrea and {Savalle}, Etienne and {Sberna}, Laura and {Sesana}, Alberto and {Shoemaker}, Deirdre and {Slutsky}, Jacob and {Sotiriou}, Thomas and {Speri}, Lorenzo and {Staab}, Martin and {Steer}, Dani{\`e}le and {Tamanini}, Nicola and {Tasinato}, Gianmassimo and {Torrado}, Jesus and {Torres-Orjuela}, Alejandro and {Toubiana}, Alexandre and {Vallisneri}, Michele and {Vecchio}, Alberto and {Volonteri}, Marta and {Yagi}, Kent and {Zwick}, Lorenz},
        title = "{LISA Definition Study Report}",
      journal = {arXiv e-prints},
         year = 2024,
        month = feb,
          eid = {arXiv:2402.07571},
        pages = {arXiv:2402.07571},
          doi = {10.48550/arXiv.2402.07571},
archivePrefix = {arXiv},
       eprint = {2402.07571},
 primaryClass = {astro-ph.CO},
       adsurl = {https://ui.adsabs.harvard.edu/abs/2024arXiv240207571C}
}

@ARTICLE{match_filter,
       author = {{Flanagan}, {\'E}anna {\'E}. and {Hughes}, Scott A.},
        title = "{Measuring gravitational waves from binary black hole coalescences. I. Signal to noise for inspiral, merger, and ringdown}",
      journal = {\prd},
         year = 1998,
        month = apr,
       volume = {57},
       number = {8},
        pages = {4535-4565},
          doi = {10.1103/PhysRevD.57.4535},
archivePrefix = {arXiv},
       eprint = {gr-qc/9701039},
 primaryClass = {gr-qc},
       adsurl = {https://ui.adsabs.harvard.edu/abs/1998PhRvD..57.4535F}
}

@ARTICLE{welch,
       author = {{Welch}, P.},
        title = "{The use of fast Fourier transform for the estimation of power spectra: A method based on time averaging over short, modified periodograms}",
      journal = {IEEE Transactions on Audio and Electroacoustics},
         year = 1967,
        month = jun,
       volume = {15},
       number = {2},
        pages = {70-73},
          doi = {10.1109/TAU.1967.1161901},
       adsurl = {https://ui.adsabs.harvard.edu/abs/1967ITAuE..15...70W}
}

@ARTICLE{noise_asd,
       author = {{Bayle}, Jean-Baptiste and {Hartwig}, Olaf},
        title = "{Unified model for the LISA measurements and instrument simulations}",
      journal = {\prd},
         year = 2023,
        month = apr,
       volume = {107},
       number = {8},
          eid = {083019},
        pages = {083019},
          doi = {10.1103/PhysRevD.107.083019},
archivePrefix = {arXiv},
       eprint = {2212.05351},
 primaryClass = {gr-qc},
       adsurl = {https://ui.adsabs.harvard.edu/abs/2023PhRvD.107h3019B}
}

@ARTICLE{gspynettree,
       author = {{{\'A}lvarez-L{\'o}pez}, Sof{\'\i}a and {Liyanage}, Annudesh and {Ding}, Julian and {Ng}, Raymond and {McIver}, Jess},
        title = "{GSpyNetTree: a signal-vs-glitch classifier for gravitational-wave event candidates}",
      journal = {Classical and Quantum Gravity},
         year = 2024,
        month = apr,
       volume = {41},
       number = {8},
          eid = {085007},
        pages = {085007},
          doi = {10.1088/1361-6382/ad2194},
archivePrefix = {arXiv},
       eprint = {2304.09977},
 primaryClass = {gr-qc},
       adsurl = {https://ui.adsabs.harvard.edu/abs/2024CQGra..41h5007A}
}

@ARTICLE{gravityspy,
       author = {{Zevin}, M. and {Coughlin}, S. and {Bahaadini}, S. and {Besler}, E. and {Rohani}, N. and {Allen}, S. and {Cabero}, M. and {Crowston}, K. and {Katsaggelos}, A.~K. and {Larson}, S.~L. and {Lee}, T.~K. and {Lintott}, C. and {Littenberg}, T.~B. and {Lundgren}, A. and {{\O}sterlund}, C. and {Smith}, J.~R. and {Trouille}, L. and {Kalogera}, V.},
        title = "{Gravity Spy: integrating advanced LIGO detector characterization, machine learning, and citizen science}",
      journal = {Classical and Quantum Gravity},
         year = 2017,
        month = mar,
       volume = {34},
       number = {6},
          eid = {064003},
        pages = {064003},
          doi = {10.1088/1361-6382/aa5cea},
archivePrefix = {arXiv},
       eprint = {1611.04596},
 primaryClass = {gr-qc},
       adsurl = {https://ui.adsabs.harvard.edu/abs/2017CQGra..34f4003Z}
}

@ARTICLE{imrphenomxhm,
       author = {{Garc{\'\i}a-Quir{\'o}s}, Cecilio and {Colleoni}, Marta and {Husa}, Sascha and {Estell{\'e}s}, H{\'e}ctor and {Pratten}, Geraint and {Ramos-Buades}, Antoni and {Mateu-Lucena}, Maite and {Jaume}, Rafel},
        title = "{Multimode frequency-domain model for the gravitational wave signal from nonprecessing black-hole binaries}",
      journal = {\prd},
         year = 2020,
        month = sep,
       volume = {102},
       number = {6},
          eid = {064002},
        pages = {064002},
          doi = {10.1103/PhysRevD.102.064002},
archivePrefix = {arXiv},
       eprint = {2001.10914},
 primaryClass = {gr-qc},
       adsurl = {https://ui.adsabs.harvard.edu/abs/2020PhRvD.102f4002G}
}

@misc{lisaorbits,
       author = {{Bayle}, Jean-Baptiste and {Hees}, Aur{\'e}lien and {Lilley}, Marc and {Le Poncin-Lafitte}, Christophe and {Martens}, Waldemar and {Joffre}, Eric},
        title = "{LISA Orbits}",
         year = 2022,
        month = feb,
          eid = {10.5281/zenodo.6412991},
          doi = {10.5281/zenodo.6412991},
      version = {2.3},
    publisher = {Zenodo},
       adsurl = {https://ui.adsabs.harvard.edu/abs/2022zndo...6412991B}
}

@ARTICLE{adamw,
       author = {{Loshchilov}, Ilya and {Hutter}, Frank},
        title = "{Decoupled Weight Decay Regularization}",
      journal = {arXiv e-prints},
         year = 2017,
        month = nov,
          eid = {arXiv:1711.05101},
        pages = {arXiv:1711.05101},
          doi = {10.48550/arXiv.1711.05101},
archivePrefix = {arXiv},
       eprint = {1711.05101},
 primaryClass = {cs.LG},
       adsurl = {https://ui.adsabs.harvard.edu/abs/2017arXiv171105101L}
}

@ARTICLE{pathfinder_overview,
       author = {{Armano}, Michele and {Audley}, Heather and {Baird}, Jonathon and {Binetruy}, Pierre and {Born}, Michael and {Bortoluzzi}, Daniele and {Castelli}, Eleanora and {Cavalleri}, Antonella and {Cesarini}, Andrea and {Cruise}, Mike and {Danzmann}, Karsten and {de Deus Silva}, Marcus and {Diepholz}, Ingo and {Dixon}, George and {Dolesi}, Rita and {Ferraioli}, Luigi and {Ferroni}, Valerio and {Fitzsimons}, Ewan and {Freschi}, Mario and {Gesa}, Luis and {Gibert}, Ferran and {Giardini}, Domenico and {Giusteri}, Roberta and {Grimani}, Catia and {Grzymisch}, Jonathan and {Harrison}, Ian and {Heinzel}, Gerhard and {Hewitson}, Martin and {Hollington}, Daniel and {Hoyland}, David and {Hueller}, Mauro and {Inchauspe}, Henri and {Jennrich}, Oliver and {Jetzer}, Philippe and {Karnesis}, Nikolaos and {Kaune}, Brigitte and {Korsakova}, Natalia and {Killow}, Christian J and {Lobo}, Alberto and {Lloro}, Ivan and {Liu}, Li and {Lopez-Zaragoza}, Juan-Pedro and {Maarschalkerweerd}, Rolf and {Mance}, Davor and {Meshksar}, Neda and {Mart{\i}n}, V and {Martin-Polo}, L and {Martino}, Joseph and {Martin-Porqueras}, Fernando and {Mateos}, Ignacio and {McNamara}, Paul and {Mendes}, Jose and {Mendes}, Luis and {Nofrarias}, Miquel and {Paczkowski}, Sarah and {Perreur-Lloyd}, Michael and {Petiteau}, Antoine and {Pivato}, Paolo and {Plagnol}, Eric and {Ramos-Castro}, Jose and {Reiche}, Jens and {Robertson}, David and {Rivas}, Francisco and {Russano}, Giuliana and {Slutsky}, Jacob and {Sopuerta}, Carlos and {Sumner}, Timothy and {Texier}, Daniel and {Thorpe}, Ira and {Vetrugno}, Daniele and {Vitale}, Stefano and {Wanner}, Gudrun and {Ward}, Harry and {Wass}, Peter and {Weber}, William and {Wissel}, Lennart and {Wittchen}, Andreas and {Zweifel}, Peter},
        title = "{LISA Pathfinder}",
      journal = {arXiv e-prints},
         year = 2019,
        month = mar,
          eid = {arXiv:1903.08924},
        pages = {arXiv:1903.08924},
          doi = {10.48550/arXiv.1903.08924},
archivePrefix = {arXiv},
       eprint = {1903.08924},
 primaryClass = {astro-ph.IM},
       adsurl = {https://ui.adsabs.harvard.edu/abs/2019arXiv190308924A}
}

@ARTICLE{pathfinder_results,
       author = {{Armano}, M. and {Audley}, H. and {Auger}, G. and {Baird}, J.~T. and {Bassan}, M. and {Binetruy}, P. and {Born}, M. and {Bortoluzzi}, D. and {Brandt}, N. and {Caleno}, M. and {Carbone}, L. and {Cavalleri}, A. and {Cesarini}, A. and {Ciani}, G. and {Congedo}, G. and {Cruise}, A.~M. and {Danzmann}, K. and {de Deus Silva}, M. and {De Rosa}, R. and {Diaz-Aguil{\'o}}, M. and {Di Fiore}, L. and {Diepholz}, I. and {Dixon}, G. and {Dolesi}, R. and {Dunbar}, N. and {Ferraioli}, L. and {Ferroni}, V. and {Fichter}, W. and {Fitzsimons}, E.~D. and {Flatscher}, R. and {Freschi}, M. and {Garc{\'\i}a Mar{\'\i}n}, A.~F. and {Garc{\'\i}a Marirrodriga}, C. and {Gerndt}, R. and {Gesa}, L. and {Gibert}, F. and {Giardini}, D. and {Giusteri}, R. and {Guzm{\'a}n}, F. and {Grado}, A. and {Grimani}, C. and {Grynagier}, A. and {Grzymisch}, J. and {Harrison}, I. and {Heinzel}, G. and {Hewitson}, M. and {Hollington}, D. and {Hoyland}, D. and {Hueller}, M. and {Inchausp{\'e}}, H. and {Jennrich}, O. and {Jetzer}, P. and {Johann}, U. and {Johlander}, B. and {Karnesis}, N. and {Kaune}, B. and {Korsakova}, N. and {Killow}, C.~J. and {Lobo}, J.~A. and {Lloro}, I. and {Liu}, L. and {L{\'o}pez-Zaragoza}, J.~P. and {Maarschalkerweerd}, R. and {Mance}, D. and {Mart{\'\i}n}, V. and {Martin-Polo}, L. and {Martino}, J. and {Martin-Porqueras}, F. and {Madden}, S. and {Mateos}, I. and {McNamara}, P.~W. and {Mendes}, J. and {Mendes}, L. and {Monsky}, A. and {Nicolodi}, D. and {Nofrarias}, M. and {Paczkowski}, S. and {Perreur-Lloyd}, M. and {Petiteau}, A. and {Pivato}, P. and {Plagnol}, E. and {Prat}, P. and {Ragnit}, U. and {Ra{\"\i}s}, B. and {Ramos-Castro}, J. and {Reiche}, J. and {Robertson}, D.~I. and {Rozemeijer}, H. and {Rivas}, F. and {Russano}, G. and {Sanju{\'a}n}, J. and {Sarra}, P. and {Schleicher}, A. and {Shaul}, D. and {Slutsky}, J. and {Sopuerta}, C.~F. and {Stanga}, R. and {Steier}, F. and {Sumner}, T. and {Texier}, D. and {Thorpe}, J.~I. and {Trenkel}, C. and {Tr{\"o}bs}, M. and {Tu}, H.~B. and {Vetrugno}, D. and {Vitale}, S. and {Wand}, V. and {Wanner}, G. and {Ward}, H. and {Warren}, C. and {Wass}, P.~J. and {Wealthy}, D. and {Weber}, W.~J. and {Wissel}, L. and {Wittchen}, A. and {Zambotti}, A. and {Zanoni}, C. and {Ziegler}, T. and {Zweifel}, P.},
        title = "{Sub-Femto-g Free Fall for Space-Based Gravitational Wave Observatories: LISA Pathfinder Results}",
      journal = {\prl},
         year = 2016,
        month = jun,
       volume = {116},
       number = {23},
          eid = {231101},
        pages = {231101},
          doi = {10.1103/PhysRevLett.116.231101},
       adsurl = {https://ui.adsabs.harvard.edu/abs/2016PhRvL.116w1101A}
}

@ARTICLE{lpf_glitches,
       author = {{Baghi}, Quentin and {Korsakova}, Natalia and {Slutsky}, Jacob and {Castelli}, Eleonora and {Karnesis}, Nikolaos and {Bayle}, Jean-Baptiste},
        title = "{Detection and characterization of instrumental transients in LISA Pathfinder and their projection to LISA}",
      journal = {\prd},
         year = 2022,
        month = feb,
       volume = {105},
       number = {4},
          eid = {042002},
        pages = {042002},
          doi = {10.1103/PhysRevD.105.042002},
archivePrefix = {arXiv},
       eprint = {2112.07490},
 primaryClass = {gr-qc},
       adsurl = {https://ui.adsabs.harvard.edu/abs/2022PhRvD.105d2002B}
}

@ARTICLE{lisa_cnn_2024,
       author = {{Houba}, Niklas and {Ferraioli}, Luigi and {Giardini}, Domenico},
        title = "{Detection and mitigation of glitches in LISA data: A machine learning approach}",
      journal = {\prd},
         year = 2024,
        month = apr,
       volume = {109},
       number = {8},
          eid = {083027},
        pages = {083027},
          doi = {10.1103/PhysRevD.109.083027},
archivePrefix = {arXiv},
       eprint = {2401.00846},
 primaryClass = {astro-ph.IM},
       adsurl = {https://ui.adsabs.harvard.edu/abs/2024PhRvD.109h3027H}
}

@ARTICLE{lisa_cnn_2025,
       author = {{Houba}, Niklas},
        title = "{Deep source separation of overlapping gravitational-wave signals and nonstationary noise artifacts}",
      journal = {Physical Review Research},
         year = 2025,
        month = aug,
       volume = {7},
       number = {3},
          eid = {033115},
        pages = {033115},
          doi = {10.1103/6bjw-xjj2},
archivePrefix = {arXiv},
       eprint = {2503.10398},
 primaryClass = {astro-ph.IM},
       adsurl = {https://ui.adsabs.harvard.edu/abs/2025PhRvR...7c3115H}
}

@ARTICLE{gwtc_4,
       author = {{The LIGO Scientific Collaboration} and {the Virgo Collaboration} and {the KAGRA Collaboration} and {Abac}, A.~G. and {Abouelfettouh}, I. and {Acernese}, F. and {Ackley}, K. and {Adamcewicz}, C. and {Adhicary}, S. and {Adhikari}, D. and {Adhikari}, N. and {Adhikari}, R.~X. and {Adkins}, V.~K. and {Afroz}, S. and {Agapito}, A. and {Agarwal}, D. and {Agathos}, M. and {Aggarwal}, N. and {Aggarwal}, S. and {Aguiar}, O.~D. and {Ahrend}, I.-L. and {Aiello}, L. and {Ain}, A. and {Ajith}, P. and {Akutsu}, T. and {Albanesi}, S. and {Ali}, W. and {Al-Kershi}, S. and {All{\'e}n{\'e}}, C. and {Allocca}, A. and {Al-Shammari}, S. and {Altin}, P.~A. and {Alvarez-Lopez}, S. and {Amar}, W. and {Amarasinghe}, O. and {Amato}, A. and {Amicucci}, F. and {Amra}, C. and {Ananyeva}, A. and {Anderson}, S.~B. and {Anderson}, W.~G. and {Andia}, M. and {Ando}, M. and {Andr{\'e}s-Carcasona}, M. and {Andri{\'c}}, T. and {Anglin}, J. and {Ansoldi}, S. and {Antelis}, J.~M. and {Antier}, S. and {Aoumi}, M. and {Appavuravther}, E.~Z. and {Appert}, S. and {Apple}, S.~K. and {Arai}, K. and {Araya}, A. and {Araya}, M.~C. and {Arca Sedda}, M. and {Areeda}, J.~S. and {Aritomi}, N. and {Armato}, F. and {Armstrong}, S. and {Arnaud}, N. and {Arogeti}, M. and {Aronson}, S.~M. and {Arun}, K.~G. and {Ashton}, G. and {Aso}, Y. and {Asprea}, L. and {Assiduo}, M. and {Assis de Souza Melo}, S. and {Aston}, S.~M. and {Astone}, P. and {Attadio}, F. and {Aubin}, F. and {AultONeal}, K. and {Avallone}, G. and {Avila}, E.~A. and {Babak}, S. and {Badger}, C. and {Bae}, S. and {Bagnasco}, S. and {Baiotti}, L. and {Bajpai}, R. and {Baka}, T. and {Baker}, A.~M. and {Baker}, K.~A. and {Baker}, T. and {Baldi}, G. and {Baldicchi}, N. and {Ball}, M. and {Ballardin}, G. and {Ballmer}, S.~W. and {Banagiri}, S. and {Banerjee}, B. and {Bankar}, D. and {Baptiste}, T.~M. and {Baral}, P. and {Baratti}, M. and {Barayoga}, J.~C. and {Barish}, B.~C. and {Barker}, D. and {Barman}, N. and {Barneo}, P. and {Barone}, F. and {Barr}, B. and {Barsotti}, L. and {Barsuglia}, M. and {Barta}, D. and {Bartoletti}, A.~M. and {Barton}, M.~A. and {Bartos}, I. and {Basalaev}, A. and {Bassiri}, R. and {Basti}, A. and {Bawaj}, M. and {Baxi}, P. and {Bayley}, J.~C. and {Baylor}, A.~C. and {Baynard}, II, P.~A. and {Bazzan}, M. and {Bedakihale}, V.~M. and {Beirnaert}, F. and {Bejger}, M. and {Belardinelli}, D. and {Bell}, A.~S. and {Bellie}, D.~S. and {Bellizzi}, L. and {Benoit}, W. and {Bentara}, I. and {Bentley}, J.~D. and {Ben Yaala}, M. and {Bera}, S. and {Bergamin}, F. and {Berger}, B.~K. and {Bernuzzi}, S. and {Beroiz}, M. and {Berry}, C.~P.~L. and {Bersanetti}, D. and {Bertheas}, T. and {Bertolini}, A. and {Betzwieser}, J. and {Beveridge}, D. and {Bevilacqua}, G. and {Bevins}, N. and {Bhandare}, R. and {Bhatt}, R. and {Bhattacharjee}, D. and {Bhattacharyya}, S. and {Bhaumik}, S. and {Biancalana}, V. and {Bianchi}, A. and {Bilenko}, I.~A. and {Billingsley}, G. and {Binetti}, A. and {Bini}, S. and {Binu}, C. and {Biot}, S. and {Birnholtz}, O. and {Biscoveanu}, S. and {Bisht}, A. and {Bitossi}, M. and {Bizouard}, M.-A. and {Blaber}, S. and {Blackburn}, J.~K. and {Blagg}, L.~A. and {Blair}, C.~D. and {Blair}, D.~G. and {Bode}, N. and {Boettner}, N. and {Boileau}, G. and {Boldrini}, M. and {Bolingbroke}, G.~N. and {Bolliand}, A. and {Bonavena}, L.~D. and {Bondarescu}, R. and {Bondu}, F. and {Bonilla}, E. and {Bonilla}, M.~S. and {Bonino}, A. and {Bonnand}, R. and {Borchers}, A. and {Borhanian}, S. and {Boschi}, V. and {Bose}, S. and {Bossilkov}, V. and {Bothra}, Y. and {Boudon}, A. and {Bourg}, L. and {Boyle}, M. and {Bozzi}, A. and {Bradaschia}, C. and {Brady}, P.~R. and {Branch}, A. and {Branchesi}, M. and {Braun}, I. and {Briant}, T. and {Brillet}, A. and {Brinkmann}, M. and {Brockill}, P. and {Brockmueller}, E.},
        title = "{GWTC-4.0: Updating the Gravitational-Wave Transient Catalog with Observations from the First Part of the Fourth LIGO-Virgo-KAGRA Observing Run}",
      journal = {arXiv e-prints},
         year = 2025,
        month = aug,
          eid = {arXiv:2508.18082},
        pages = {arXiv:2508.18082},
          doi = {10.48550/arXiv.2508.18082},
archivePrefix = {arXiv},
       eprint = {2508.18082},
 primaryClass = {gr-qc},
       adsurl = {https://ui.adsabs.harvard.edu/abs/2025arXiv250818082T}
}

@ARTICLE{ligo_bbh,
       author = {{Abbott}, B.~P. and {Abbott}, R. and {Abbott}, T.~D. and {Abernathy}, M.~R. and {Acernese}, F. and {Ackley}, K. and {Adams}, C. and {Adams}, T. and {Addesso}, P. and {Adhikari}, R.~X. and {Adya}, V.~B. and {Affeldt}, C. and {Agathos}, M. and {Agatsuma}, K. and {Aggarwal}, N. and {Aguiar}, O.~D. and {Aiello}, L. and {Ain}, A. and {Ajith}, P. and {Allen}, B. and {Allocca}, A. and {Altin}, P.~A. and {Anderson}, S.~B. and {Anderson}, W.~G. and {Arai}, K. and {Arain}, M.~A. and {Araya}, M.~C. and {Arceneaux}, C.~C. and {Areeda}, J.~S. and {Arnaud}, N. and {Arun}, K.~G. and {Ascenzi}, S. and {Ashton}, G. and {Ast}, M. and {Aston}, S.~M. and {Astone}, P. and {Aufmuth}, P. and {Aulbert}, C. and {Babak}, S. and {Bacon}, P. and {Bader}, M.~K.~M. and {Baker}, P.~T. and {Baldaccini}, F. and {Ballardin}, G. and {Ballmer}, S.~W. and {Barayoga}, J.~C. and {Barclay}, S.~E. and {Barish}, B.~C. and {Barker}, D. and {Barone}, F. and {Barr}, B. and {Barsotti}, L. and {Barsuglia}, M. and {Barta}, D. and {Bartlett}, J. and {Barton}, M.~A. and {Bartos}, I. and {Bassiri}, R. and {Basti}, A. and {Batch}, J.~C. and {Baune}, C. and {Bavigadda}, V. and {Bazzan}, M. and {Behnke}, B. and {Bejger}, M. and {Belczynski}, C. and {Bell}, A.~S. and {Bell}, C.~J. and {Berger}, B.~K. and {Bergman}, J. and {Bergmann}, G. and {Berry}, C.~P.~L. and {Bersanetti}, D. and {Bertolini}, A. and {Betzwieser}, J. and {Bhagwat}, S. and {Bhandare}, R. and {Bilenko}, I.~A. and {Billingsley}, G. and {Birch}, J. and {Birney}, I.~A. and {Birnholtz}, O. and {Biscans}, S. and {Bisht}, A. and {Bitossi}, M. and {Biwer}, C. and {Bizouard}, M.~A. and {Blackburn}, J.~K. and {Blair}, C.~D. and {Blair}, D.~G. and {Blair}, R.~M. and {Bloemen}, S. and {Bock}, O. and {Bodiya}, T.~P. and {Boer}, M. and {Bogaert}, G. and {Bogan}, C. and {Bohe}, A. and {Bojtos}, P. and {Bond}, C. and {Bondu}, F. and {Bonnand}, R. and {Boom}, B.~A. and {Bork}, R. and {Boschi}, V. and {Bose}, S. and {Bouffanais}, Y. and {Bozzi}, A. and {Bradaschia}, C. and {Brady}, P.~R. and {Braginsky}, V.~B. and {Branchesi}, M. and {Brau}, J.~E. and {Briant}, T. and {Brillet}, A. and {Brinkmann}, M. and {Brisson}, V. and {Brockill}, P. and {Brooks}, A.~F. and {Brown}, D.~A. and {Brown}, D.~D. and {Brown}, N.~M. and {Buchanan}, C.~C. and {Buikema}, A. and {Bulik}, T. and {Bulten}, H.~J. and {Buonanno}, A. and {Buskulic}, D. and {Buy}, C. and {Byer}, R.~L. and {Cabero}, M. and {Cadonati}, L. and {Cagnoli}, G. and {Cahillane}, C. and {Bustillo}, J. Calder{\'o}n and {Callister}, T. and {Calloni}, E. and {Camp}, J.~B. and {Cannon}, K.~C. and {Cao}, J. and {Capano}, C.~D. and {Capocasa}, E. and {Carbognani}, F. and {Caride}, S. and {Diaz}, J. Casanueva and {Casentini}, C. and {Caudill}, S. and {Cavagli{\`a}}, M. and {Cavalier}, F. and {Cavalieri}, R. and {Cella}, G. and {Cepeda}, C.~B. and {Baiardi}, L. Cerboni and {Cerretani}, G. and {Cesarini}, E. and {Chakraborty}, R. and {Chalermsongsak}, T. and {Chamberlin}, S.~J. and {Chan}, M. and {Chao}, S. and {Charlton}, P. and {Chassande-Mottin}, E. and {Chen}, H.~Y. and {Chen}, Y. and {Cheng}, C. and {Chincarini}, A. and {Chiummo}, A. and {Cho}, H.~S. and {Cho}, M. and {Chow}, J.~H. and {Christensen}, N. and {Chu}, Q. and {Chua}, S. and {Chung}, S. and {Ciani}, G. and {Clara}, F. and {Clark}, J.~A. and {Cleva}, F. and {Coccia}, E. and {Cohadon}, P.-F. and {Colla}, A. and {Collette}, C.~G. and {Cominsky}, L. and {Constancio}, M. and {Conte}, A. and {Conti}, L. and {Cook}, D. and {Corbitt}, T.~R. and {Cornish}, N. and {Corsi}, A. and {Cortese}, S. and {Costa}, C.~A. and {Coughlin}, M.~W. and {Coughlin}, S.~B. and {Coulon}, J.-P. and {Countryman}, S.~T. and {Couvares}, P. and {Cowan}, E.~E. and {Coward}, D.~M. and {Cowart}, M.~J.},
        title = "{Observation of Gravitational Waves from a Binary Black Hole Merger}",
      journal = {\prl},
         year = 2016,
        month = feb,
       volume = {116},
       number = {6},
          eid = {061102},
        pages = {061102},
          doi = {10.1103/PhysRevLett.116.061102},
archivePrefix = {arXiv},
       eprint = {1602.03837},
 primaryClass = {gr-qc},
       adsurl = {https://ui.adsabs.harvard.edu/abs/2016PhRvL.116f1102A}
}

@ARTICLE{lisa_gr,
       author = {{Gair}, Jonathan R. and {Vallisneri}, Michele and {Larson}, Shane L. and {Baker}, John G.},
        title = "{Testing General Relativity with Low-Frequency, Space-Based Gravitational-Wave Detectors}",
      journal = {Living Reviews in Relativity},
         year = 2013,
        month = dec,
       volume = {16},
       number = {1},
          eid = {7},
        pages = {7},
          doi = {10.12942/lrr-2013-7},
archivePrefix = {arXiv},
       eprint = {1212.5575},
 primaryClass = {gr-qc},
       adsurl = {https://ui.adsabs.harvard.edu/abs/2013LRR....16....7G}
}

@ARTICLE{lisa_smbhb,
       author = {{Mangiagli}, Alberto and {Caprini}, Chiara and {Volonteri}, Marta and {Marsat}, Sylvain and {Vergani}, Susanna and {Tamanini}, Nicola and {Inchausp{\'e}}, Henri},
        title = "{Massive black hole binaries in LISA: Multimessenger prospects and electromagnetic counterparts}",
      journal = {\prd},
         year = 2022,
        month = nov,
       volume = {106},
       number = {10},
          eid = {103017},
        pages = {103017},
          doi = {10.1103/PhysRevD.106.103017},
archivePrefix = {arXiv},
       eprint = {2207.10678},
 primaryClass = {astro-ph.HE},
       adsurl = {https://ui.adsabs.harvard.edu/abs/2022PhRvD.106j3017M}
}

@ARTICLE{lisa_bwd,
       author = {{Littenberg}, T.~B. and {Larson}, S.~L. and {Nelemans}, G. and {Cornish}, N.~J.},
        title = "{Prospects for observing ultracompact binaries with space-based gravitational wave interferometers and optical telescopes}",
      journal = {\mnras},
         year = 2013,
        month = mar,
       volume = {429},
       number = {3},
        pages = {2361-2365},
          doi = {10.1093/mnras/sts507},
archivePrefix = {arXiv},
       eprint = {1207.4848},
 primaryClass = {astro-ph.IM},
       adsurl = {https://ui.adsabs.harvard.edu/abs/2013MNRAS.429.2361L}
}

@ARTICLE{lisa_bwd2,
       author = {{Korol}, Valeriya and {Rossi}, Elena M. and {Groot}, Paul J. and {Nelemans}, Gijs and {Toonen}, Silvia and {Brown}, Anthony G.~A.},
        title = "{Prospects for detection of detached double white dwarf binaries with Gaia, LSST and LISA}",
      journal = {\mnras},
         year = 2017,
        month = sep,
       volume = {470},
       number = {2},
        pages = {1894-1910},
          doi = {10.1093/mnras/stx1285},
archivePrefix = {arXiv},
       eprint = {1703.02555},
 primaryClass = {astro-ph.HE},
       adsurl = {https://ui.adsabs.harvard.edu/abs/2017MNRAS.470.1894K}
}

@ARTICLE{lisa_ht,
       author = {{Hoang}, Bao-Minh and {Naoz}, Smadar and {Kocsis}, Bence and {Farr}, Will M. and {McIver}, Jessica},
        title = "{Detecting Supermassive Black Hole-induced Binary Eccentricity Oscillations with LISA}",
      journal = {\apjl},
         year = 2019,
        month = apr,
       volume = {875},
       number = {2},
          eid = {L31},
        pages = {L31},
          doi = {10.3847/2041-8213/ab14f7},
archivePrefix = {arXiv},
       eprint = {1903.00134},
 primaryClass = {astro-ph.HE},
       adsurl = {https://ui.adsabs.harvard.edu/abs/2019ApJ...875L..31H}
}

@ARTICLE{ligo_glitch,
       author = {{Davis}, D. and {Littenberg}, T.~B. and {Romero-Shaw}, I.~M. and {Millhouse}, M. and {McIver}, J. and {Di Renzo}, F. and {Ashton}, G.},
        title = "{Subtracting glitches from gravitational-wave detector data during the third LIGO-Virgo observing run}",
      journal = {Classical and Quantum Gravity},
         year = 2022,
        month = dec,
       volume = {39},
       number = {24},
          eid = {245013},
        pages = {245013},
          doi = {10.1088/1361-6382/aca238},
archivePrefix = {arXiv},
       eprint = {2207.03429},
 primaryClass = {astro-ph.IM},
       adsurl = {https://ui.adsabs.harvard.edu/abs/2022CQGra..39x5013D}
}

@ARTICLE{alan_search,
       author = {{Knee}, Alan M. and {McIver}, Jess and {Naoz}, Smadar and {Romero-Shaw}, Isobel M. and {Hoang}, Bao-Minh and {Grishin}, Evgeni},
        title = "{Detecting Gravitational-wave Bursts from Black Hole Binaries in the Galactic Center with LISA}",
      journal = {\apjl},
         year = 2024,
        month = aug,
       volume = {971},
       number = {2},
          eid = {L38},
        pages = {L38},
          doi = {10.3847/2041-8213/ad6a10},
archivePrefix = {arXiv},
       eprint = {2404.12571},
 primaryClass = {astro-ph.HE},
       adsurl = {https://ui.adsabs.harvard.edu/abs/2024ApJ...971L..38K}
}

@ARTICLE{scipy,
       author = {{Virtanen}, Pauli and {Gommers}, Ralf and {Oliphant}, Travis E. and {Haberland}, Matt and {Reddy}, Tyler and {Cournapeau}, David and {Burovski}, Evgeni and {Peterson}, Pearu and {Weckesser}, Warren and {Bright}, Jonathan and {van der Walt}, St{\'e}fan J. and {Brett}, Matthew and {Wilson}, Joshua and {Millman}, K. Jarrod and {Mayorov}, Nikolay and {Nelson}, Andrew R.~J. and {Jones}, Eric and {Kern}, Robert and {Larson}, Eric and {Carey}, C.~J. and {Polat}, {\.I}lhan and {Feng}, Yu and {Moore}, Eric W. and {VanderPlas}, Jake and {Laxalde}, Denis and {Perktold}, Josef and {Cimrman}, Robert and {Henriksen}, Ian and {Quintero}, E.~A. and {Harris}, Charles R. and {Archibald}, Anne M. and {Ribeiro}, Ant{\^o}nio H. and {Pedregosa}, Fabian and {van Mulbregt}, Paul and {SciPy 1.  0 Contributors}},
        title = "{SciPy 1.0: fundamental algorithms for scientific computing in Python}",
      journal = {Nature Medicine},
         year = 2020,
        month = feb,
       volume = {17},
        pages = {261-272},
          doi = {10.1038/s41592-019-0686-2},
archivePrefix = {arXiv},
       eprint = {1907.10121},
 primaryClass = {cs.MS},
       adsurl = {https://ui.adsabs.harvard.edu/abs/2020NaMet..17..261V}
}

@misc{pytdi,
       author = {{Staab}, Martin and {Bayle}, Jean-Baptiste and {Hartwig}, Olaf},
        title = "{PyTDI}",
         year = 2023,
        month = oct,
          eid = {10.5281/zenodo.6351736},
          doi = {10.5281/zenodo.6351736},
      version = {1.3.1},
    publisher = {Zenodo},
       adsurl = {https://ui.adsabs.harvard.edu/abs/2022zndo...6351736S}
}

@ARTICLE{pytorch,
       author = {{Paszke}, Adam and {Gross}, Sam and {Massa}, Francisco and {Lerer}, Adam and {Bradbury}, James and {Chanan}, Gregory and {Killeen}, Trevor and {Lin}, Zeming and {Gimelshein}, Natalia and {Antiga}, Luca and {Desmaison}, Alban and {K{\"o}pf}, Andreas and {Yang}, Edward and {DeVito}, Zach and {Raison}, Martin and {Tejani}, Alykhan and {Chilamkurthy}, Sasank and {Steiner}, Benoit and {Fang}, Lu and {Bai}, Junjie and {Chintala}, Soumith},
        title = "{PyTorch: An Imperative Style, High-Performance Deep Learning Library}",
      journal = {arXiv e-prints},
         year = 2019,
        month = dec,
          eid = {arXiv:1912.01703},
        pages = {arXiv:1912.01703},
          doi = {10.48550/arXiv.1912.01703},
archivePrefix = {arXiv},
       eprint = {1912.01703},
 primaryClass = {cs.LG},
       adsurl = {https://ui.adsabs.harvard.edu/abs/2019arXiv191201703P}
}

@ARTICLE{random_forest,
       author = {{Breiman}, Leo},
        title = "{Random Forests.}",
      journal = {Machine Learning},
         year = 2001,
        month = jan,
       volume = {45},
        pages = {5-32},
          doi = {10.1023/A:1010933404324},
       adsurl = {https://ui.adsabs.harvard.edu/abs/2001MachL..45....5B}
}

@ARTICLE{gwtc_1,
       author = {{Abbott}, B.~P. and {Abbott}, R. and {Abbott}, T.~D. and {Abraham}, S. and {Acernese}, F. and {Ackley}, K. and {Adams}, C. and {Adhikari}, R.~X. and {Adya}, V.~B. and {Affeldt}, C. and {Agathos}, M. and {Agatsuma}, K. and {Aggarwal}, N. and {Aguiar}, O.~D. and {Aiello}, L. and {Ain}, A. and {Ajith}, P. and {Allen}, G. and {Allocca}, A. and {Aloy}, M.~A. and {Altin}, P.~A. and {Amato}, A. and {Ananyeva}, A. and {Anderson}, S.~B. and {Anderson}, W.~G. and {Angelova}, S.~V. and {Antier}, S. and {Appert}, S. and {Arai}, K. and {Araya}, M.~C. and {Areeda}, J.~S. and {Ar{\`e}ne}, M. and {Arnaud}, N. and {Arun}, K.~G. and {Ascenzi}, S. and {Ashton}, G. and {Aston}, S.~M. and {Astone}, P. and {Aubin}, F. and {Aufmuth}, P. and {AultONeal}, K. and {Austin}, C. and {Avendano}, V. and {Avila-Alvarez}, A. and {Babak}, S. and {Bacon}, P. and {Badaracco}, F. and {Bader}, M.~K.~M. and {Bae}, S. and {Baker}, P.~T. and {Baldaccini}, F. and {Ballardin}, G. and {Ballmer}, S.~W. and {Banagiri}, S. and {Barayoga}, J.~C. and {Barclay}, S.~E. and {Barish}, B.~C. and {Barker}, D. and {Barkett}, K. and {Barnum}, S. and {Barone}, F. and {Barr}, B. and {Barsotti}, L. and {Barsuglia}, M. and {Barta}, D. and {Bartlett}, J. and {Bartos}, I. and {Bassiri}, R. and {Basti}, A. and {Bawaj}, M. and {Bayley}, J.~C. and {Bazzan}, M. and {B{\'e}csy}, B. and {Bejger}, M. and {Belahcene}, I. and {Bell}, A.~S. and {Beniwal}, D. and {Berger}, B.~K. and {Bergmann}, G. and {Bernuzzi}, S. and {Bero}, J.~J. and {Berry}, C.~P.~L. and {Bersanetti}, D. and {Bertolini}, A. and {Betzwieser}, J. and {Bhandare}, R. and {Bidler}, J. and {Bilenko}, I.~A. and {Bilgili}, S.~A. and {Billingsley}, G. and {Birch}, J. and {Birney}, R. and {Birnholtz}, O. and {Biscans}, S. and {Biscoveanu}, S. and {Bisht}, A. and {Bitossi}, M. and {Bizouard}, M.~A. and {Blackburn}, J.~K. and {Blackman}, J. and {Blair}, C.~D. and {Blair}, D.~G. and {Blair}, R.~M. and {Bloemen}, S. and {Bode}, N. and {Boer}, M. and {Boetzel}, Y. and {Bogaert}, G. and {Bondu}, F. and {Bonilla}, E. and {Bonnand}, R. and {Booker}, P. and {Boom}, B.~A. and {Booth}, C.~D. and {Bork}, R. and {Boschi}, V. and {Bose}, S. and {Bossie}, K. and {Bossilkov}, V. and {Bosveld}, J. and {Bouffanais}, Y. and {Bozzi}, A. and {Bradaschia}, C. and {Brady}, P.~R. and {Bramley}, A. and {Branchesi}, M. and {Brau}, J.~E. and {Briant}, T. and {Briggs}, J.~H. and {Brighenti}, F. and {Brillet}, A. and {Brinkmann}, M. and {Brisson}, V. and {Brockill}, P. and {Brooks}, A.~F. and {Brown}, D.~D. and {Brunett}, S. and {Buikema}, A. and {Bulik}, T. and {Bulten}, H.~J. and {Buonanno}, A. and {Buskulic}, D. and {Bustamante Rosell}, M.~J. and {Buy}, C. and {Byer}, R.~L. and {Cabero}, M. and {Cadonati}, L. and {Cagnoli}, G. and {Cahillane}, C. and {Calder{\'o}n Bustillo}, J. and {Callister}, T.~A. and {Calloni}, E. and {Camp}, J.~B. and {Campbell}, W.~A. and {Canepa}, M. and {Cannon}, K.~C. and {Cao}, H. and {Cao}, J. and {Capocasa}, E. and {Carbognani}, F. and {Caride}, S. and {Carney}, M.~F. and {Carullo}, G. and {Casanueva Diaz}, J. and {Casentini}, C. and {Caudill}, S. and {Cavagli{\`a}}, M. and {Cavalier}, F. and {Cavalieri}, R. and {Cella}, G. and {Cerd{\'a}-Dur{\'a}n}, P. and {Cerretani}, G. and {Cesarini}, E. and {Chaibi}, O. and {Chakravarti}, K. and {Chamberlin}, S.~J. and {Chan}, M. and {Chao}, S. and {Charlton}, P. and {Chase}, E.~A. and {Chassande-Mottin}, E. and {Chatterjee}, D. and {Chaturvedi}, M. and {Chatziioannou}, K. and {Cheeseboro}, B.~D. and {Chen}, H.~Y. and {Chen}, X. and {Chen}, Y. and {Cheng}, H.-P. and {Cheong}, C.~K. and {Chia}, H.~Y. and {Chincarini}, A. and {Chiummo}, A. and {Cho}, G. and {Cho}, H.~S. and {Cho}, M. and {Christensen}, N. and {Chu}, Q. and {Chua}, S. and {Chung}, K.~W.},
        title = "{GWTC-1: A Gravitational-Wave Transient Catalog of Compact Binary Mergers Observed by LIGO and Virgo during the First and Second Observing Runs}",
      journal = {Physical Review X},
         year = 2019,
        month = jul,
       volume = {9},
       number = {3},
          eid = {031040},
        pages = {031040},
          doi = {10.1103/PhysRevX.9.031040},
archivePrefix = {arXiv},
       eprint = {1811.12907},
 primaryClass = {astro-ph.HE},
       adsurl = {https://ui.adsabs.harvard.edu/abs/2019PhRvX...9c1040A}
}

@ARTICLE{gwtc_2,
       author = {{Abbott}, R. and {Abbott}, T.~D. and {Abraham}, S. and {Acernese}, F. and {Ackley}, K. and {Adams}, A. and {Adams}, C. and {Adhikari}, R.~X. and {Adya}, V.~B. and {Affeldt}, C. and {Agathos}, M. and {Agatsuma}, K. and {Aggarwal}, N. and {Aguiar}, O.~D. and {Aiello}, L. and {Ain}, A. and {Ajith}, P. and {Akcay}, S. and {Allen}, G. and {Allocca}, A. and {Altin}, P.~A. and {Amato}, A. and {Anand}, S. and {Ananyeva}, A. and {Anderson}, S.~B. and {Anderson}, W.~G. and {Angelova}, S.~V. and {Ansoldi}, S. and {Antelis}, J.~M. and {Antier}, S. and {Appert}, S. and {Arai}, K. and {Araya}, M.~C. and {Areeda}, J.~S. and {Ar{\`e}ne}, M. and {Arnaud}, N. and {Aronson}, S.~M. and {Arun}, K.~G. and {Asali}, Y. and {Ascenzi}, S. and {Ashton}, G. and {Aston}, S.~M. and {Astone}, P. and {Aubin}, F. and {Aufmuth}, P. and {AultONeal}, K. and {Austin}, C. and {Avendano}, V. and {Babak}, S. and {Badaracco}, F. and {Bader}, M.~K.~M. and {Bae}, S. and {Baer}, A.~M. and {Bagnasco}, S. and {Baird}, J. and {Ball}, M. and {Ballardin}, G. and {Ballmer}, S.~W. and {Bals}, A. and {Balsamo}, A. and {Baltus}, G. and {Banagiri}, S. and {Bankar}, D. and {Bankar}, R.~S. and {Barayoga}, J.~C. and {Barbieri}, C. and {Barish}, B.~C. and {Barker}, D. and {Barneo}, P. and {Barnum}, S. and {Barone}, F. and {Barr}, B. and {Barsotti}, L. and {Barsuglia}, M. and {Barta}, D. and {Bartlett}, J. and {Bartos}, I. and {Bassiri}, R. and {Basti}, A. and {Bawaj}, M. and {Bayley}, J.~C. and {Bazzan}, M. and {Becher}, B.~R. and {B{\'e}csy}, B. and {Bedakihale}, V.~M. and {Bejger}, M. and {Belahcene}, I. and {Beniwal}, D. and {Benjamin}, M.~G. and {Bennett}, T.~F. and {Bentley}, J.~D. and {Bergamin}, F. and {Berger}, B.~K. and {Bergmann}, G. and {Bernuzzi}, S. and {Berry}, C.~P.~L. and {Bersanetti}, D. and {Bertolini}, A. and {Betzwieser}, J. and {Bhandare}, R. and {Bhandari}, A.~V. and {Bhattacharjee}, D. and {Bidler}, J. and {Bilenko}, I.~A. and {Billingsley}, G. and {Birney}, R. and {Birnholtz}, O. and {Biscans}, S. and {Bischi}, M. and {Biscoveanu}, S. and {Bisht}, A. and {Bitossi}, M. and {Bizouard}, M.-A. and {Blackburn}, J.~K. and {Blackman}, J. and {Blair}, C.~D. and {Blair}, D.~G. and {Blair}, R.~M. and {Blanch}, O. and {Bobba}, F. and {Bode}, N. and {Boer}, M. and {Boetzel}, Y. and {Bogaert}, G. and {Boldrini}, M. and {Bondu}, F. and {Bonilla}, E. and {Bonnand}, R. and {Booker}, P. and {Boom}, B.~A. and {Bork}, R. and {Boschi}, V. and {Bose}, S. and {Bossilkov}, V. and {Boudart}, V. and {Bouffanais}, Y. and {Bozzi}, A. and {Bradaschia}, C. and {Brady}, P.~R. and {Bramley}, A. and {Branchesi}, M. and {Brau}, J.~E. and {Breschi}, M. and {Briant}, T. and {Briggs}, J.~H. and {Brighenti}, F. and {Brillet}, A. and {Brinkmann}, M. and {Brockill}, P. and {Brooks}, A.~F. and {Brooks}, J. and {Brown}, D.~D. and {Brunett}, S. and {Bruno}, G. and {Bruntz}, R. and {Buikema}, A. and {Bulik}, T. and {Bulten}, H.~J. and {Buonanno}, A. and {Buscicchio}, R. and {Buskulic}, D. and {Byer}, R.~L. and {Cabero}, M. and {Cadonati}, L. and {Caesar}, M. and {Cagnoli}, G. and {Cahillane}, C. and {Calder{\'o}n Bustillo}, J. and {Callaghan}, J.~D. and {Callister}, T.~A. and {Calloni}, E. and {Camp}, J.~B. and {Canepa}, M. and {Cannon}, K.~C. and {Cao}, H. and {Cao}, J. and {Carapella}, G. and {Carbognani}, F. and {Carney}, M.~F. and {Carpinelli}, M. and {Carullo}, G. and {Carver}, T.~L. and {Casanueva Diaz}, J. and {Casentini}, C. and {Caudill}, S. and {Cavagli{\`a}}, M. and {Cavalier}, F. and {Cavalieri}, R. and {Cella}, G. and {Cerd{\'a}-Dur{\'a}n}, P. and {Cesarini}, E. and {Chaibi}, W. and {Chakravarti}, K. and {Chan}, C.-L. and {Chan}, C. and {Chandra}, K. and {Chanial}, P. and {Chao}, S. and {Charlton}, P. and {Chase}, E.~A.},
        title = "{GWTC-2: Compact Binary Coalescences Observed by LIGO and Virgo during the First Half of the Third Observing Run}",
      journal = {Physical Review X},
         year = 2021,
        month = apr,
       volume = {11},
       number = {2},
          eid = {021053},
        pages = {021053},
          doi = {10.1103/PhysRevX.11.021053},
archivePrefix = {arXiv},
       eprint = {2010.14527},
 primaryClass = {gr-qc},
       adsurl = {https://ui.adsabs.harvard.edu/abs/2021PhRvX..11b1053A}
}

@ARTICLE{gwtc_3,
       author = {{Abbott}, R. and {Abbott}, T.~D. and {Acernese}, F. and {Ackley}, K. and {Adams}, C. and {Adhikari}, N. and {Adhikari}, R.~X. and {Adya}, V.~B. and {Affeldt}, C. and {Agarwal}, D. and {Agathos}, M. and {Agatsuma}, K. and {Aggarwal}, N. and {Aguiar}, O.~D. and {Aiello}, L. and {Ain}, A. and {Ajith}, P. and {Akcay}, S. and {Akutsu}, T. and {Albanesi}, S. and {Allocca}, A. and {Altin}, P.~A. and {Amato}, A. and {Anand}, C. and {Anand}, S. and {Ananyeva}, A. and {Anderson}, S.~B. and {Anderson}, W.~G. and {Ando}, M. and {Andrade}, T. and {Andres}, N. and {Andri{\'c}}, T. and {Angelova}, S.~V. and {Ansoldi}, S. and {Antelis}, J.~M. and {Antier}, S. and {Appert}, S. and {Arai}, Koji and {Arai}, Koya and {Arai}, Y. and {Araki}, S. and {Araya}, A. and {Araya}, M.~C. and {Areeda}, J.~S. and {Ar{\`e}ne}, M. and {Aritomi}, N. and {Arnaud}, N. and {Arogeti}, M. and {Aronson}, S.~M. and {Arun}, K.~G. and {Asada}, H. and {Asali}, Y. and {Ashton}, G. and {Aso}, Y. and {Assiduo}, M. and {Aston}, S.~M. and {Astone}, P. and {Aubin}, F. and {Austin}, C. and {Babak}, S. and {Badaracco}, F. and {Bader}, M.~K.~M. and {Badger}, C. and {Bae}, S. and {Bae}, Y. and {Baer}, A.~M. and {Bagnasco}, S. and {Bai}, Y. and {Baiotti}, L. and {Baird}, J. and {Bajpai}, R. and {Ball}, M. and {Ballardin}, G. and {Ballmer}, S.~W. and {Balsamo}, A. and {Baltus}, G. and {Banagiri}, S. and {Bankar}, D. and {Barayoga}, J.~C. and {Barbieri}, C. and {Barish}, B.~C. and {Barker}, D. and {Barneo}, P. and {Barone}, F. and {Barr}, B. and {Barsotti}, L. and {Barsuglia}, M. and {Barta}, D. and {Bartlett}, J. and {Barton}, M.~A. and {Bartos}, I. and {Bassiri}, R. and {Basti}, A. and {Bawaj}, M. and {Bayley}, J.~C. and {Baylor}, A.~C. and {Bazzan}, M. and {B{\'e}csy}, B. and {Bedakihale}, V.~M. and {Bejger}, M. and {Belahcene}, I. and {Benedetto}, V. and {Beniwal}, D. and {Bennett}, T.~F. and {Bentley}, J.~D. and {Benyaala}, M. and {Bergamin}, F. and {Berger}, B.~K. and {Bernuzzi}, S. and {Berry}, C.~P.~L. and {Bersanetti}, D. and {Bertolini}, A. and {Betzwieser}, J. and {Beveridge}, D. and {Bhandare}, R. and {Bhardwaj}, U. and {Bhattacharjee}, D. and {Bhaumik}, S. and {Bilenko}, I.~A. and {Billingsley}, G. and {Bini}, S. and {Birney}, R. and {Birnholtz}, O. and {Biscans}, S. and {Bischi}, M. and {Biscoveanu}, S. and {Bisht}, A. and {Biswas}, B. and {Bitossi}, M. and {Bizouard}, M.-A. and {Blackburn}, J.~K. and {Blair}, C.~D. and {Blair}, D.~G. and {Blair}, R.~M. and {Bobba}, F. and {Bode}, N. and {Boer}, M. and {Bogaert}, G. and {Boldrini}, M. and {Bonavena}, L.~D. and {Bondu}, F. and {Bonilla}, E. and {Bonnand}, R. and {Booker}, P. and {Boom}, B.~A. and {Bork}, R. and {Boschi}, V. and {Bose}, N. and {Bose}, S. and {Bossilkov}, V. and {Boudart}, V. and {Bouffanais}, Y. and {Bozzi}, A. and {Bradaschia}, C. and {Brady}, P.~R. and {Bramley}, A. and {Branch}, A. and {Branchesi}, M. and {Brandt}, J. and {Brau}, J.~E. and {Breschi}, M. and {Briant}, T. and {Briggs}, J.~H. and {Brillet}, A. and {Brinkmann}, M. and {Brockill}, P. and {Brooks}, A.~F. and {Brooks}, J. and {Brown}, D.~D. and {Brunett}, S. and {Bruno}, G. and {Bruntz}, R. and {Bryant}, J. and {Bulik}, T. and {Bulten}, H.~J. and {Buonanno}, A. and {Buscicchio}, R. and {Buskulic}, D. and {Buy}, C. and {Byer}, R.~L. and {Davies}, G.~S. Cabourn and {Cadonati}, L. and {Cagnoli}, G. and {Cahillane}, C. and {Bustillo}, J. Calder{\'o}n and {Callaghan}, J.~D. and {Callister}, T.~A. and {Calloni}, E. and {Cameron}, J. and {Camp}, J.~B. and {Canepa}, M. and {Canevarolo}, S. and {Cannavacciuolo}, M. and {Cannon}, K.~C. and {Cao}, H. and {Cao}, Z. and {Capocasa}, E. and {Capote}, E. and {Carapella}, G. and {Carbognani}, F.},
        title = "{GWTC-3: Compact Binary Coalescences Observed by LIGO and Virgo during the Second Part of the Third Observing Run}",
      journal = {Physical Review X},
         year = 2023,
        month = oct,
       volume = {13},
       number = {4},
          eid = {041039},
        pages = {041039},
          doi = {10.1103/PhysRevX.13.041039},
archivePrefix = {arXiv},
       eprint = {2111.03606},
 primaryClass = {gr-qc},
       adsurl = {https://ui.adsabs.harvard.edu/abs/2023PhRvX..13d1039A}
}

@ARTICLE{q_transform,
       author = {{Chatterji}, S. and {Blackburn}, L. and {Martin}, G. and {Katsavounidis}, E.},
        title = "{Multiresolution techniques for the detection of gravitational-wave bursts}",
      journal = {Classical and Quantum Gravity},
         year = 2004,
        month = oct,
       volume = {21},
       number = {20},
        pages = {S1809-S1818},
          doi = {10.1088/0264-9381/21/20/024},
archivePrefix = {arXiv},
       eprint = {gr-qc/0412119},
 primaryClass = {gr-qc},
       adsurl = {https://ui.adsabs.harvard.edu/abs/2004CQGra..21S1809C}
}

@ARTICLE{scott_rule,
       author = {{Scott}, David W.},
        title = "{On optimal and data-based histograms}",
      journal = {Biometrics},
         year = 1979,
        month = jan,
       volume = {66},
       number = {3},
        pages = {605-610},
          doi = {10.1093/biomet/66.3.605},
       adsurl = {https://ui.adsabs.harvard.edu/abs/1979Biome..66..605S}
}

@ARTICLE{bcewithlogits,
       author = {{Terven}, Juan and {Cordova-Esparza}, Diana M. and {Ramirez-Pedraza}, Alfonso and {Chavez-Urbiola}, Edgar A. and {Romero-Gonzalez}, Julio A.},
        title = "{Loss Functions and Metrics in Deep Learning}",
      journal = {arXiv e-prints},
         year = 2023,
        month = jul,
          eid = {arXiv:2307.02694},
        pages = {arXiv:2307.02694},
          doi = {10.48550/arXiv.2307.02694},
archivePrefix = {arXiv},
       eprint = {2307.02694},
 primaryClass = {stat.ML},
       adsurl = {https://ui.adsabs.harvard.edu/abs/2023arXiv230702694T}
}
\bibliographystyle{aasjournal}
\end{document}